# Persistent currents in normal metal rings: comparing high-precision experiment with theory

A. C. Bleszynski-Jayich,[1] W. E. Shanks,[1] B. Peaudecerf,[1] E. Ginossar,[1] F. von Oppen,[2] L. Glazman,[1,3] and J. G. E. Harris[1,3]

[1] *Department of Physics, Yale University, New Haven CT, 06520 USA*
[2] *Institut für Theoretische Physik, Freie Universität Berlin, Fachbereich Physik, 14195 Berlin, Germany*
[3] *Department of Applied Physics, Yale University, New Haven CT, 06520 USA*

Quantum mechanics predicts that the equilibrium state of a resistive electrical circuit contains a dissipationless current. This persistent current has been the focus of considerable theoretical and experimental work, but its basic properties remain a topic of controversy. The main experimental challenges in studying persistent currents have been the small signals they produce and their exceptional sensitivity to their environment. To address these issues we have developed a new technique for detecting persistent currents which offers greatly improved sensitivity and reduced measurement back action. This allows us to measure the persistent current in metal rings over a wider range of temperature, ring size, and magnetic field than has been possible previously. We find that measurements of both a single ring and arrays of rings agree well with calculations based on a model of non-interacting electrons.

An electrical current induced in a resistive circuit will rapidly decay in the absence of an applied voltage. This decay reflects the tendency of the circuit's electrons to dissipate energy and relax to their ground state. However quantum mechanics predicts that the electrons' many-body ground state (and, at finite temperature, their thermal equilibrium state) may itself contain a "persistent" current which flows through the resistive circuit without dissipating energy or decaying. A dissipationless equilibrium current flowing through a resistive circuit is highly counterintuitive, but it has a familiar analog in atomic physics: some atomic species' electronic ground states possess non-zero orbital angular momentum, equivalent to a current circulating around the atom.

Theoretical treatments of persistent currents (PC) in resistive metal rings have been developed over a number of decades (see [*1,2*] and references therein). Calculations which take into account the electrons' inevitable coupling to a static disorder potential and a fluctuating thermal bath predict several general features. A micron-diameter ring will support a PC of $I \sim 1$ nA at temperatures $T \lesssim 1$ K. A magnetic flux $\Phi$ threading the ring will break time-reversal symmetry, allowing the PC to flow in a particular direction around the ring. Furthermore, the Aharonov-Bohm effect will require $I$ to be periodic in $\Phi$ with period $\Phi_0 = h/e$, thereby providing a clear-cut experimental signature of the PC.

These predictions have attracted considerable interest, but measuring the PC is challenging for a number of reasons. For example, the PC flows only within the ring and so cannot be measured using a conventional ammeter. Experiments to date[2,3] have mostly used SQUIDs to infer the PC from the magnetic field it produces. Interpretation of these measurements has been complicated by the SQUIDs' low signal-to-noise ratio (SNR) and the uncontrolled back action of the SQUID's ac Josephson oscillations, which may drive non-equilibrium currents in the rings. In addition, SQUIDs perform optimally in low magnetic fields; this limits the maximum $\Phi$ which can be applied to the rings, allowing observation of only a few oscillations of $I(\Phi)$ and complicating the subtraction of background signals unrelated to the PC.

Experiments to date have produced a number of confusing results in apparent contradiction with theory and even amongst the experiments themselves[2,3]. These conflicts have remained without a clear resolution for nearly twenty years, suggesting that our understanding of how to measure and/or calculate the ground state properties of as simple a system as an isolated metal ring may be incomplete.

More recent theoretical work has predicted that the PC is highly sensitive to a variety of subtle effects, including electron-electron interactions[4,5,6,7], the ring's coupling to its electromagnetic environment[8], and trace magnetic impurities within the ring[9]. These theories have not explained the experimental results to date, but they do indicate that accurate measurements of the PC would be able to address a number of interesting questions in many-body condensed matter physics (in addition to resolving the long-standing controversy described above).

Here we present measurements of the PC in resistive metal rings using a micromechanical detector with orders of magnitude greater sensitivity and lower back-action than SQUID-based detectors. This approach allows us to measure the PC in a single ring and arrays of rings as a function of ring size, temperature, and the magnitude and orientation of the magnetic field over a much broader range than has been possible previously. We find quantitative agreement between these measurements and calculations based on a model of diffusive, non-interacting electrons. This agreement is supported by independent measurements of the rings' electrical properties.

Figures 1A-C show single-crystal Si cantilevers with integrated Al rings (their fabrication is described elsewhere[*10*]). All the PC measurements were made in magnetic fields well above the critical field of Al, ensuring the rings are in their normal (rather than superconducting) state. The parameters of the four ring samples measured here are given in Table 1.

In the presence of a magnetic field $\vec{B}$, each ring's current $I$ produces a torque on the cantilever $\vec{\tau} = \vec{\mu} \times \vec{B}$ as well as a shift $\delta\nu$ in the cantilever's resonant frequency $\nu$. Here $\vec{\mu} = \pi r^2 I \hat{n}$ is the magnetic moment of the PC, $r$ is the ring radius, and $\hat{n}$ is the unit vector normal to the ring. We infer $I(B)$ from measurements of $\delta\nu(B)$; the conversion between $\delta\nu(B)$ and $I(B)$ is described in the Supporting Online Material (SOM).

To monitor $\nu$ we drive the cantilever in a phase-locked loop. The cantilever is driven via a piezoelectric element, and the cantilever's displacement is monitored by a fiber-optic interferometer[*11*]. The cantilever's thermally limited force sensitivity is ~ 2.9 aN/Hz$^{1/2}$ at $T =$ 300 mK, corresponding to a magnetic moment sensitivity of $\sim 11$ $\mu_B/Hz^{1/2}$ and a current sensitivity of $\sim 20$ pA/Hz$^{1/2}$ for a ring with $r = 400$ nm at $B = 8$ T. By comparison, SQUID magnetometers achieve a current sensitivity $\gtrsim 5$ nA/Hz$^{1/2}$ for a similar ring[12,13,14]. We have shown previously that a cantilever's noise temperature and the electron temperature of a metal sample at the end of a cantilever equilibrate with the fridge temperature for the conditions used here[11].

The frequency shift of a cantilever containing an array of $N = 1680$ nominally identical rings with $r = 308$ nm at $T = 323$ mK is shown in Fig. 1D as a function of $B$. Oscillations with a period $\approx 20$ mT, corresponding to a flux $h/e$ through each ring, are visible in the raw data. Depending upon $r$ and $\theta$ (the angle between $\vec{B}$ and the plane of the ring) we observe as many as 450 oscillations over a 5.5 T range of $B$ (data shown in the SOM).

Figure 1E shows the data from Fig. 1D after subtracting the smooth background and converting the data from $\delta\nu(B)$ to $I(B)$ using the expressions in the SOM. The left-hand axis in Fig. 1E shows $I_\Sigma$, the total PC inferred from the measurement, which is the sum of the PC from each ring in the array. The right-hand axis shows the estimated typical single-ring PC: $I_{typ} = I_\Sigma/\sqrt{N}$. This relationship between $I_{typ}$ and $I_\Sigma$ arises because the PC in each ring is predicted to oscillate as a function of $B$ with a phase which depends upon the ring's microscopic disorder, and thus is assumed to be random from ring to ring. This assumption is verified below.

To establish that $\delta\nu$ provides a reliable measure of the PC we measured $I_{typ}(B)$ as a function of several experimental conditions: the laser power incident on the cantilever, the amplitude and frequency of the cantilever's motion, the polarity and orientation of the magnetic field, and the presence or absence of room temperature electronics connected to the cryostat. These data are shown in the SOM, and indicate that the measurements of $I_{typ}(B)$ are independent of these parameters (for the conditions of our experiment) and reflect the equilibrium persistent current in the rings.

Figures 2A-C show $I_{typ}(B)$ for arrays of rings with three different radii: $r$ = 308 nm, 418 nm, and 793 nm. We have also measured a single ring with $r$ = 418 nm, shown in Fig. 2D. Figures 2E-H show $\left|\tilde{I}_{typ}(f_\Phi)\right|$, the absolute value of the Fourier transform of the data in Figs 2A-D ($f_\Phi$ is the "flux frequency" in units of $(h/e)^{-1}$). Figures 2I-L show $G_{typ}(\delta B)$, the autocorrelation of $I_{typ}(B)$ for each of these samples. $G_{typ}(\delta B)$ is calculated from measurements of $I_{typ}(B)$ taken over a much broader range of $B$ than is shown in Figs. 2A-D; the complete data is shown in the SOM.

We can draw a number of conclusions from a qualitative examination of this data. First, we note that $I_{typ}(B)$ oscillates with a period $\approx h/e$, but also contains an aperiodic modulation which broadens the peaks in $\tilde{I}_{typ}(f_\Phi)$ and causes $G_{typ}(\delta B)$ to decay at large $\delta B$. This modulation is due to the fact that we apply a uniform $B$ to the sample, leading to magnetic flux inside the metal of each ring given by $\Phi_M = BA_M$ where $A_M$ is the area of the metal projected along $\bar{B}$. This leads to a new effective disorder potential (and hence a randomization of the phase of the $I(B)$ oscillations) each time $\Phi_M$ changes by $\sim \Phi_0$[15]. As a result, the peaks in $\tilde{I}_{typ}(f_\Phi)$ span a band of $f_\Phi$ roughly bounded by the rings' inner and outer radii (the blue bars in Figs. 2E-H), and the decay of $G_{typ}(\delta B)$ is found to occur on a field scale (defined as the half-width half-max of $G_{typ}(\delta B)$ [16]) $B_c = \kappa\Phi_0/A_M$. Here $\kappa$ is a constant which is predicted[17] to be $\approx 1$; we find $1 < \kappa < 3$ in these samples. For the array samples, ring-to-ring variations in $r$ (estimated to be $\sim 1\%$) should contribute negligibly to $B_c$ and the peak widths in $\tilde{I}_{typ}(f_\Phi)$. The fact that the $r$ = 418 nm array and the $r$ = 418 nm single ring show similar peak width and $B_c$ indicate that variations in $r$ do not affect the signal appreciably.

It is clear from Fig. 2 that the PC is smaller in larger rings. This is consistent with the prediction[18] that the typical amplitude $I_{h/e}(T = 0)$ of the $h/e$-periodic Fourier component of $I(\Phi)$ at $T = 0$ corresponds roughly to the current produced by a single electron diffusing around the ring at the Fermi energy, and hence should scale as $1/r^2$. In addition, $I_{h/e}(T)$ is predicted[18] to decrease on a temperature scale (known as the Thouless temperature) $T_T \propto 1/r^2$ corresponding to the scale of disorder-induced correlations in the ring's spectrum of single-electron states.

In Fig. 2E a small peak at $f_\Phi = 2$ can be seen, corresponding to the second harmonic of $I(\Phi)$. This harmonic has attracted particular attention because under some conditions it has a component which is not random from ring to ring[4,19,20]. The signal from such a non-random, "average" current would scale as $I_\Sigma^{(avg)} \propto N$ rather than $\sqrt{N}$. Furthermore, the amplitude of $I_\Sigma^{(avg)}$ can be strongly enhanced by electron-electron interactions[4] and other effects[8,9]. However $I_\Sigma^{(avg)}$ arises from the cooperon contribution to the PC and so requires time-reversal symmetry within the metal, which in our experiments is broken by $\Phi_M$. We calculate that $\Phi_M$ suppresses $I_\Sigma^{(avg)}$ by a factor $\sim e^{-2\pi r/1.3\ell_B}$ (where the magnetic length $\ell_B = \sqrt{h/eB}$), which for this experiment should render $I_\Sigma^{(avg)}$ unobservably small. As a result, the peak in Fig. 2E at $f_\Phi = 2$ presumably reflects the random component of the second harmonic of $I(\Phi)$, which is predicted[18] to have a zero-temperature amplitude $I_{h/2e}(0) = I_{h/e}(0)/2^{3/2}$, to be suppressed on a temperature scale = $T_T/4$, and to produce a signal with the same $\sqrt{N}$ scaling as $I_{h/e}$.

We now turn to a more quantitative analysis of the data. Theory predicts[18] that, for each independent realization of the disorder potential, $I_{h/pe}$ (the $p^{th}$ harmonic of $I(\Phi)$) is drawn randomly from a distribution with a mean $\langle I_{h/pe} \rangle_D = 0$ and an rms value $\langle I_{h/pe}^2 \rangle_D^{1/2}$ which in general is non-zero. Here $\langle \cdots \rangle_D$ represents an average over disorder potentials. The quantity $\langle I_{h/pe}^2 \rangle_D^{1/2}$ can be calculated explicitly as a function of $r$, $T$, $p$, and the electrons' diffusion constant $D$ for a variety of models.

To compare our data against these calculations, we make use of the fact that $\langle I_{h/pe}^2 \rangle_D^{1/2}$ can be extracted from a measurement of $I_\Sigma(B)$ when the measurement record spans many $B_c$. When this condition is satisfied, averages performed with respect to $B$ are equivalent to averages performed with respect to disorder realizations, and it is straightforward to show that the area under a peak in $|\tilde{I}_{typ}(f_\Phi)|^2$ (cf. Figs. 2E-H) at $f_\Phi = p$ is simply related to $\langle I_{h/pe}^2 \rangle_D^{1/2}$:

$$\left[ \int_{f_\Phi^-}^{f_\Phi^+} \left( |\tilde{I}_{typ}(f_\Phi)|^2 - b(f_\Phi) \right) df_\Phi \right]^{1/2} = \langle I_{h/pe}^2 \rangle_D^{1/2}. \tag{1}$$

Here $b$ is the noise floor in $|\tilde{I}_{typ}(f_\Phi)|^2$, and is estimated from the portions of the data away from the peaks. We take the limits of integration $f_\Phi^+$ and $f_\Phi^-$ to be roughly the values of $f_\Phi$ corresponding to $h/pe$ flux periodicity through the outer and inner radii of the ring, respectively. In previous experiments, $\langle I_{h/pe}^2 \rangle_D$ could only be determined from successive measurements of individual, nominally identical rings[21,3]. This approach was limited by the low SNR achieved in single-ring measurements and practical limits on the number of nominally identical rings ($\approx 15$) which could be measured.

Measurements of $\langle I_{h/e}^2 \rangle_D^{1/2}$ for each sample and $\langle I_{h/2e}^2 \rangle_D^{1/2}$ for the smallest rings are shown in Fig. 3 as a function of $T$ for $\theta = 45°$ (open symbols) and $\theta = 6°$ (closed symbols). From Fig. 3 it can be seen that the PC in larger rings decays more quickly with $T$ than in smaller rings, and that $\langle I^2_{h/2e}\rangle_D^{1/2}$ decays more quickly than $\langle I^2_{h/e}\rangle_D^{1/2}$, consistent with the discussion above. In addition, the agreement between the data for the $r = 418$ nm array and the $r = 418$ nm single ring indicates that the PC signal scales as $\sqrt{N}$ and hence that the PC is random from ring to ring.

The solid lines in Fig. 3 are fits to theoretical predictions in which $\langle I^2_{h/pe}\rangle_D^{1/2}$ is calculated for diffusive noninteracting electrons. This calculation closely follows that of Ref. [18] but takes into account the presence of the large magnetic field $B$ inside the metal (which lifts the spin degeneracy and removes the cooperon contribution to the PC) as well as spin-orbit scattering (the rings' circumference exceeds the spin-orbit scattering length, as discussed in the SOM). We find:

$$\langle I^2_{h/pe}(T)\rangle_D = g\left(p^2 \frac{T}{T_T}\right)\langle I^2_{h/pe}(0)\rangle_D \qquad (2)$$

where $g(x) = \frac{\pi^6}{3}x^2 \sum_{n=1}^{\infty} n\exp[-(2\pi^3 nx)^{1/2}]$, $\langle I^2_{h/pe}(0)\rangle_D^{1/2} = 0.37 p^{-3/2} \frac{3eD}{(2\pi r)^2}$, and $T_T = \frac{\hbar \pi^2 D}{k_B (2\pi r)^2}$.

The data from each sample in Fig. 3 was fit separately, in each case using $D$ as the only fitting parameter. The best fit values of $D$ are listed in Table 1. These values are typical for high-purity evaporated aluminum wires of the dimensions used here[22,23]; however, to further constrain the comparison between our data and theory we also independently determined $D$ from the resistivity of a co-deposited wire (the wire's properties are listed in Table I). This measurement is described in detail in the SOM and provides a value of $D$ in good agreement with the values extracted from the persistent current measurements. We note that the values of $D$ in Table 1 show a correlation with the samples' linewidths which may reflect the increased contribution of surface scattering in the narrower samples.

The calculation leading to Eq. 2 assumes the phase coherent motion of free electrons around the ring. Measurements of the phase coherence length $L_\Phi(T)$ in the co-deposited wire are described in the SOM, and show that $L_\Phi \gg 2\pi r$ for nearly all the temperatures at which the PC is observable. The closest approach between $L_\Phi$ and $2\pi r$ at a temperature where the PC can still be observed occurs in the 308 nm array at $T = 3$ K where we find $L_\Phi(3\ \mathrm{K}) = 1.86 \times (2\pi r)$. It is conceivable that the more rapid decrease in $\langle I^2_{h/e} \rangle_D^{1/2}$ observed in this sample above $T = 2$ K (Fig. 3) is due to dephasing; however it is not possible to test this hypothesis in the other samples, as the larger rings' PC is well below the noise floor when $L_\Phi(T) = 1.86 \times (2\pi r)$. To the best of our knowledge the effect of dephasing upon the PC has not been calculated.

In conclusion, we have measured the persistent current in normal metal rings over a wide range of temperature, ring size, array size, magnetic field magnitude, and magnetic field orientation with high signal-to-noise ratio, excellent background rejection, and low measurement back-action. These measurements indicate that the rings' equilibrium state is well-described by the diffusive non-interacting electron model. In addition to providing a clear experimental picture of persistent currents in simple metallic rings, these results open the possibility of using measurements of the PC to search for ultra-low temperature phase transitions[6], or to study a variety of many-body and environmental effects relevant to quantum phase transitions and quantum coherence in solid state qubits[24,25]. Furthermore, the micromechanical detectors used here are well-suited to studying the PC in circuits driven out of equilibrium (e.g., by the controlled introduction of microwave radiation)[8]. The properties of persistent currents in these regimes have received relatively little attention to date but could offer new insights into the behavior of isolated nanoelectronic systems.[26]

| Sample | $r$ (nm) | $w$ (nm) | $d$ (nm) | $N$ | $D$ (cm$^2$/s) |
|---|---|---|---|---|---|
| 308 nm array | 308 | 115 | 90 | 1680 | 271 ± 2.6 |
| 418 nm array | 418 | 85 | 90 | 990 | 214 ± 3.3 |
| 793 nm array | 793 | 85 | 90 | 242 | 205 ± 6.5 |
| 418 nm ring | 418 | 85 | 90 | 1 | 215 ± 4.6 |
| Wire (see SOM) | 289,000 (length) | 115 | 90 | 1 | 260 ± 12 |

**Table 1.** Sample parameters. For each of the four ring samples, the rings' mean radius $r$, linewidth $w$, and thickness $d$ are listed, along with the number $N$ of rings in the sample. The electrons' diffusion constant $D$, extracted from the fits in Fig. 3, is given. The stated errors are statistical errors in the fits. An additional 6% error in $D$ is estimated for uncertainties in the overall calibration, as discussed in the SOM. The fifth sample is the co-deposited wire used in the transport measurements described in the SOM. For this sample $D$ was determined from the wire's resistivity.

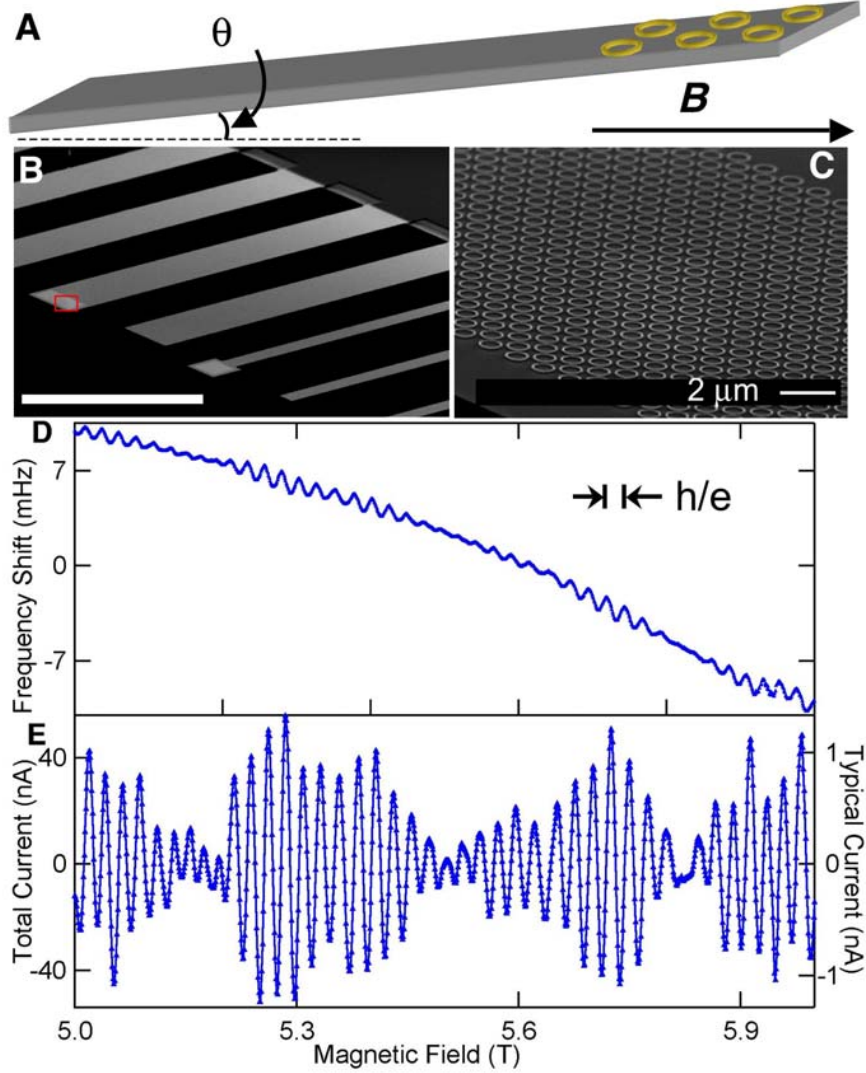

**Fig. 1.** (**A**) Cantilever torque magnetometry schematic. An array of metal rings is integrated onto the end of a cantilever. The cantilever is mounted in a $^3$He refrigerator. A magnetic field $B$ is applied at an angle $\theta$ from the plane of the rings. The out-of-plane component of $B$ provides magnetic flux $\Phi$ through the ring. The in-plane component of $B$ exerts a torque on the rings' magnetic moment and causes a shift in the cantilever's resonant frequency $\delta\nu$. Laser interferometry is used to monitor the cantilever's motion and to determine $\delta\nu$. (**B**) A scanning electron micrograph of several Si cantilevers similar to those used in the experiment. The light regions at the end of some of the cantilevers are arrays of Al rings. The scale bar is 100 μm. The individual rings are visible in (**C**), which shows a magnified view of the region in (**B**) outlined in red. (**D**) Raw data showing $\delta\nu$ as a function of $B$ for an array of $N = 1680$ rings with $r = 308$ nm at $T = 365$ mK and $\theta = 45°$. (**E**) Persistent current inferred from the frequency shift data in (**D**) after subtracting a smooth background from the raw data. The left-hand axis shows the total current $I_\Sigma$ in the array and the right-hand axis shows the estimated typical per-ring current $I_{\text{typ}} = I_\Sigma/\sqrt{N}$. Oscillations with a characteristic period of ~ 20 mT (corresponding to $\Phi = h/e$) are visible in (**D**) and (**E**).

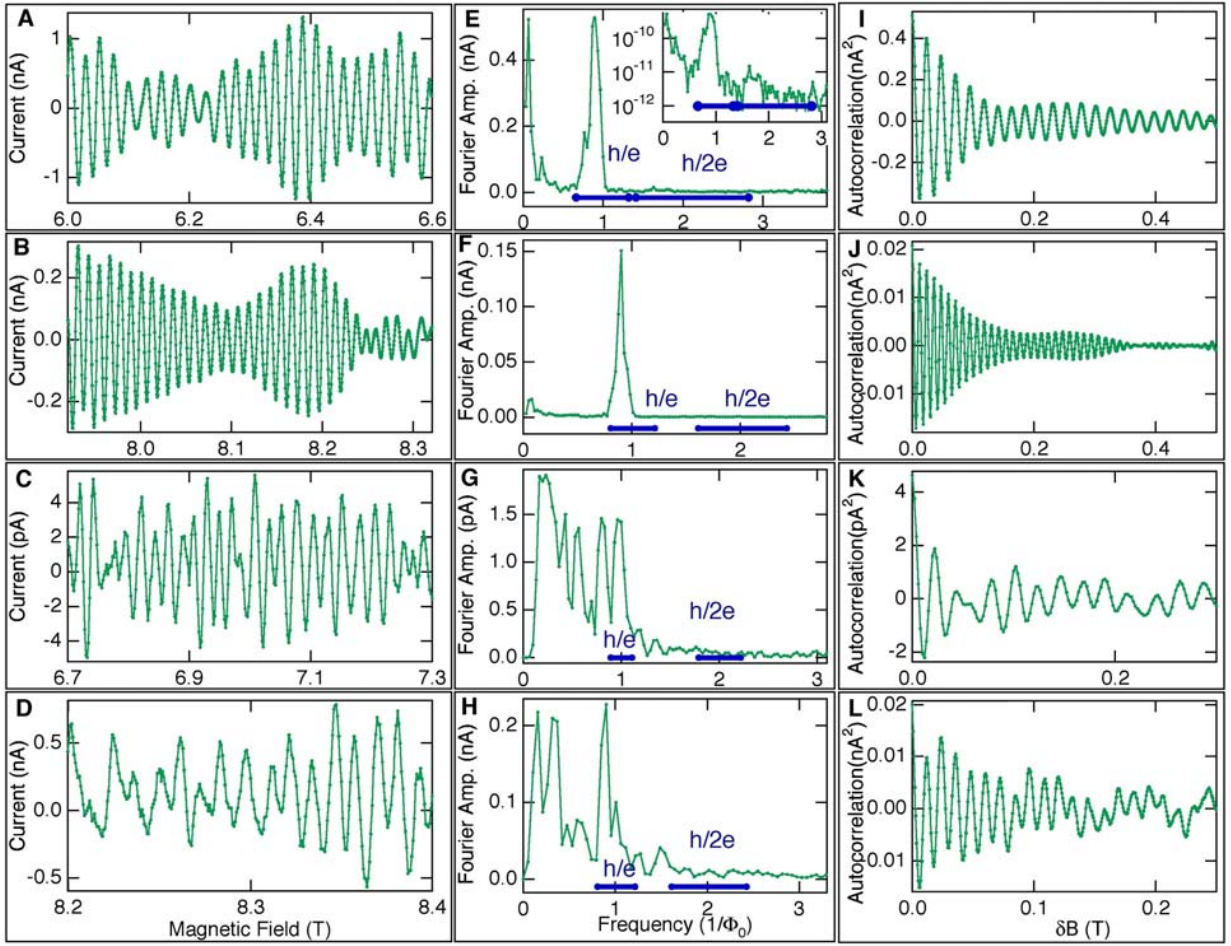

**Fig. 2.** Persistent current versus magnetic field in: (**A**) the 308 nm array for $T = 365$ mK and $\theta = 45°$, (**B**) the 418 nm array for $T = 365$ mK, $\theta = 45°$, (**C**) the 793 nm array for $T = 323$ mK, $\theta = 6°$, and (**D**) the 418 nm ring for 365 mK, $\theta = 45°$. In each case a smooth background has been removed. (**E**)-(**H**) show Fourier transforms of the data in (**A**)-(**D**). The expected $h/e$ and $h/2e$ periodicities are indicated by the blue bar. The bars' widths reflect the rings' linewidth $w$. A small $h/2e$ peak is present in (**E**) (visible in the log-scale graph, inset). (**I**)-(**L**) show the autocorrelation functions of the data shown in (**A**)-(**D**), but computed over a field range $\Delta B$ larger than shown in (**A**)-(**D**): $\Delta B =$ (**I**) 5.4 T, (**J**) 5.3 T, (**K**) 0.6 T, and (**L**) 1.1 T (full data shown in SOM).

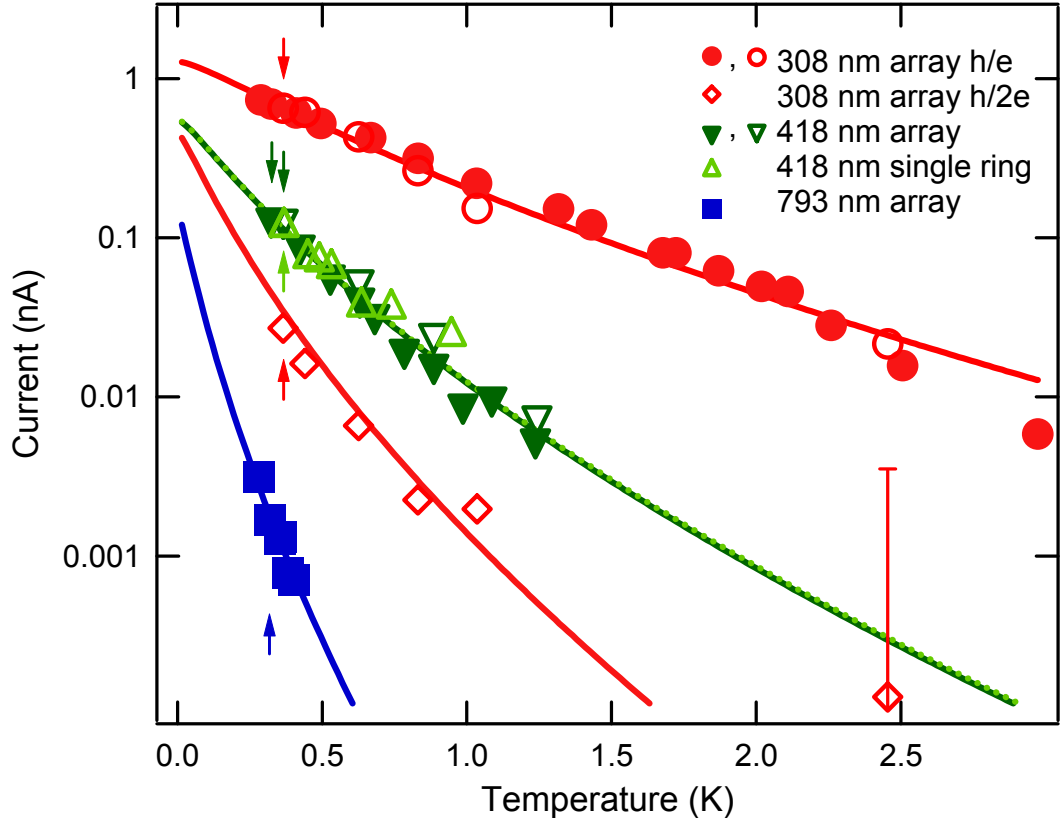

**Fig. 3.** Temperature dependence of the *h/e* and *h/2e* Fourier components of the current per ring. The vertical axis indicates $\langle I_{h/e}^2 \rangle_D^{1/2}$ and $\langle I_{h/2e}^2 \rangle_D^{1/2}$, the rms values of the Fourier amplitudes of the persistent current. In each data set, the solid points were taken with $\theta = 45°$ while for the hollow points $\theta = 6°$. The arrows indicate the data points derived from $I(B)$ measurements taken over a magnetic field range much greater than $B_c$; other data points are derived from the scaling of $I(B)$ measured over a smaller range of $B$. The lines (solid for array samples, dotted for the single ring) are fits to the prediction for noninteracting diffusive electrons. The electron diffusion constant $D$ is the only fitting parameter, and is listed in Table I.

[26] We thank M. Devoret, R. Ilic, T. Ojanen, and J. C. Sankey for their assistance. A.C.B.-J., W.E.S. and J.G.E.H. are supported by NSF grants 0706380 and 0653377. F.v.O. is supported in part by DIP. J.G.E.H. acknowledges support from the Sloan Foundation. A.C.B.-J. acknowledges support from UNESCO-L'Oreal. L.G. is supported in part by DOE grant DE-FG02-08ER46482. F.v.O and L.G. acknowledge the hospitality of KITP in the final stages of this work.

# Supporting Online Material

"Persistent currents in normal metal rings: comparing high-precision experiment with theory" by Bleszynski-Jayich *et al*.

## 1. Transport measurements

Transport measurements were performed on an aluminum wire codeposited with the rings studied in this report. The wire was deposited adjacent to the cantilevers on the same silicon wafer (approximately 5 mm away) and had a length of 289 µm, a linewidth $w$ of 115 ± 6 nm (measured by SEM) and a thickness $d$ of 90 ± 1 nm (measured by AFM). An SEM image of a similar wire is shown in Fig. S1. All measurements were performed with the wire in a three terminal ac resistance bridge using a lock-in amplifier[1,2]. Excitation currents were chosen to be low enough that the measurement results were independent of the magnitude of excitation. In order to prevent high frequency noise from reaching the sample, the measurement lines incorporated coaxial low pass filters at the room temperature feedthrough into the cryostat (3 dB frequency ~ 1.9 MHz) and at the cold stage of the fridge (3 dB frequency ~ 80 MHz). For the magnetoresistance measurements described below, the chip was mounted with the magnetic field normal to its surface.

### 1.1 Resistivity measurement

The wire's resistivity ρ was obtained by measuring the total change in resistance of the sample at 360 mK as the magnetic field was swept through the wire's superconducting critical field (Fig. S2). The diffusion constant $D$ was then calculated using the Einstein relation

$$\rho^{-1} = e^2 g D \qquad (S1)$$

with $e$ the electron charge and $g$ the electron density of states per unit volume at the Fermi level. The density of states can be written in terms of the free electron density $n$ and the Fermi energy $\varepsilon_F$ as $g = 3n/2\varepsilon_F$. With $n$ = 1.81 × 10$^{29}$ m$^{-3}$ and $\varepsilon_F$ =11.5 eV for aluminum[3], the wire's measured resistivity of $\rho$ = 1.03 ± 0.05 × 10$^{-8}$ Ωm corresponds to a diffusion constant of $D$ = 0.026 ± 0.01 m$^2$/s. To avoid confusion in the following sections, we denote this value of $D$ as $D_\rho$.

## 1.2 Superconducting critical field measurement

The wire's superconducting critical field $H_C$ was measured as a function of temperature $T$ near the wire's superconducting transition temperature $T_C$. In the Ginzburg Landau framework valid for a dirty superconductor near $T_C$, the superconducting critical field for a thin wire lying on a plane normal to the applied magnetic field can be written as

$$H_C(T) = \frac{\sqrt{12}h}{\pi e w \sqrt{D}} \sqrt{k_B(T_C - T)} \qquad (S2)$$

where $h$ is Planck's constant and $k_B$ is the Boltzmann constant[4]. The superconducting critical field was measured by sweeping the magnetic field at different sample temperatures and observing the change in resistance from the normal to the superconducting state. The superconducting critical field was taken to be the field at which the wire resistance reached a fixed fraction of the normal state resistance. In Fig. S3, the measured $H_C(T)$ and a fit to Eq. S2 are shown. The extracted fit parameters are $T_C = 1.19$ K and $D = 122 \pm 5$ cm$^2$/s. We denote this value of $D$ as $D_{Hc}$.

We note that Eq. S2 is applicable only when the electrons' elastic scattering length $\ell_e$ is much smaller than the wire's transverse dimensions $w$ and $t$. If we use the value of $D_\rho$ determined from the resistivity measurements described above and a Fermi velocity $v_F = 2.0 \times 10^6$ m/s, then we find a value of $\ell_e = 3 D_\rho/v_F = 40$ nm. This indicates that $\ell_e \sim w, d$, and hence that Eq. S2 is not valid. As a result we do not consider $D_{Hc}$ to provide an accurate estimate of $D$, and include it here only for completeness. Discrepancies between $D_{Hc}$ and $D_\rho$ due to the breakdown of Eq. S2 have been noted previously[5].

## 1.3 Measurement of the electron phase coherence and spin orbit lengths

The electron phase coherence and spin orbit scattering lengths were extracted from measurements of the magnetoresistance of the same wire at temperatures above $T_c$. Such measurements were first performed in aluminum wires two decades ago and have been reviewed previously[6].

The coherent interference of time reversed trajectories leads to an increase in the probability for a quasiparticle to return to its original position and thus an increase in electrical resistance, a phenomenon known as weak localization. The presence of a magnetic field suppresses weak localization by breaking time reversal symmetry and allows a direct measure of electron phase coherence through the resulting magnetoresistance. Spin orbit scattering can also modify the spin components of time reversed paths and thus the weak localization contribution to conductivity. The analytic form for the weak localization correction to the resistance $R$ in a magnetic field $B$ is given by

$$\frac{\delta R^{WL}}{R} = \frac{R(B)-R(B=0)}{R(B=0)} = \frac{3}{2}f_1\left(B,b\left(L_\varphi+\frac{4}{3}L_{SO}\right)\right) - \frac{1}{2}f_1\left(B,b(L_\varphi)\right) \qquad (S3)$$

where $L_\varphi$ is electron phase coherence length and $L_{SO}$ is the spin orbit length. The function $f(B,B_1)$ is given by

$$f(B,B_1) = R_\square \frac{e^2}{\pi\hbar}\left(\frac{b(w)}{B_1}\right)^{1/2}\left(1+\frac{B^2}{48b(w)B_1}\right)^{-1/2} \qquad (S4)$$

with $R_\square = \rho/d$ the sheet resistance per square unit of the wire. The field scale $b(w)$ is given by

$$b(l) = \frac{\hbar}{4el^2} \qquad (S5)$$

where $l$ is in units of length. This form for the weak localization correction to the magnetoresistance is derived from a perturbative calculation and is valid for $B < 12b(w) \sim 300$ mT for our wire.[6]

Just above $T_C$, superconducting fluctuations result in a small, temperature dependent population of Cooper pairs, which reduce the resistance of the metal. Further above $T_C$, in the temperature range relevant to our measurements, Cooper pairs from superconducting fluctuations are too short-lived to contribute directly to the conductivity. However, after a Cooper pair decays, the two electron quasiparticle wave functions are still correlated and provide a contribution to the conductivity like a Cooper pair, known as the Maki Thompson contribution, as long as the electrons maintain phase coherence. Because all Cooper pairs are composed of electrons in the singlet state, spin orbit scattering does not affect the Maki Thompson contribution to the conductivity. The correlation between quasiparticles formed by the decay of a Cooper pair has a theoretically similar description to the cooperon which describes weak localization and thus both effects have similar analytic forms for their contributions to the magnetoresistance. Specifically, the Maki Thompson correction to the resistance obeys:

$$\frac{\delta R^{MT}}{R} = \frac{R(B) - R(B=0)}{R(B=0)} = -\beta\left(\frac{T}{T_C}\right) f\left(B, b(L_\varphi)\right) \quad \text{(S6)}$$

where $\beta(t)$ is a function introduced by Larkin[7] which diverges logarithmically as $t \to 1$. Eq. S6 is valid provided that $\frac{\hbar D}{L_\varphi^2} \ll k_B T \ln(T/T_C)$ and $B \ll \frac{k_B T}{4De} \ln(T/T_C)$.[8]

Magnetoresistance measurements were made at a series of wire temperatures above $T_c$ between 1.8 and 10 K. Because of the limitations on the validity of Eq. S6, measurements could only be made at relatively high temperatures compared to those for which we measured persistent currents. Fig. S4 shows magnetoresistance measurements with fits to the sum of Eqs. S3 and S6 for three different temperatures. From these fits we determine that spin-orbit scattering contributes significantly to the magnetoresistance only at higher temperatures where the Maki Thompson contribution is small.

For temperatures above 4 K, the magnetoresistance data were fit with both $L_\varphi$ and $L_{SO}$ as fitting parameters. In this range, $L_{SO}$ was measured to be $1.1 \pm 0.25$ μm and observed to be independent of temperature. Following this analysis, the data for the whole temperature range was fit with $L_{SO}$ fixed to 1.1 μm and $L_\varphi$ as the only free parameter. In Fig. S5, the fitted values for $L_\varphi$ found in this way are plotted versus temperature.

We note that $L_{SO}$ is less than the circumference of the rings used in the persistent current measurements. For this reason, we have fit our data to the expression for the persistent current (Eq. 2 of the main paper) which assumes strong spin-orbit coupling.

The electron phase coherence length is limited to a finite value due to scattering processes in which electrons change energy. For the temperature regime of our measurements, the processes expected to be dominant are electron-phonon and electron-electron scattering. Scattering from magnetic impurities should be negligible for our high purity aluminum film (total impurity concentration 10 ± 5 ppm; Fe concentration 0.5 ppm.). The electron-phonon phase scattering rate $\tau_{ep}^{-1}$ follows the form

$$\tau_{ep}^{-1} = A_{ep} T^3 \qquad (S7)$$

where $A_{ep} = 9.1 \times 10^6 \, \text{s}^{-1} \text{K}^{-3}$ for bulk aluminum[6,9]. For the conditions of our measurement, the electron-electron phase scattering rate $\tau_{ee}^{-1}$ has been observed to be dominated by the process of multiple collisions with small energy transfers and to follow the form[10]

$$\tau_{ee}^{-1} = A_{ee} T^{2/3} = \left( \frac{R_\square e^2}{4\hbar} \left( \frac{k_B}{\hbar} \right) \frac{\sqrt{D}}{w} \right)^{2/3} T^{2/3}. \qquad (S8)$$

The total electron phase breaking rate $\tau_\varphi^{-1}$ is approximately equal to the sum of these two rates[10]. The blue lines in Fig. S5 show a fit to the measured $L_\varphi(T)$ using Eqs. S7 and S8 and the relation $L_\varphi = \sqrt{D \tau_\varphi}$ with $A_{ep}$ and $D$ as the free parameters. The fitted values are $A_{ep} = (1.21 \pm 0.07) \times 10^7 \, \text{s}^{-1} \text{K}^{-3}$ and $D = 700 \pm 30$ cm²/s. We denote this value of $D$ as $D_\varphi$. While $D_\varphi$ differs from $D_\rho$ and the value of $D$ extracted from the persistent current, we note that $L_\varphi$ only depends upon $D^{1/3}$ and so provides a relatively weak constraint on $D$. In addition we note that Eq. S8 is derived without considering the presence of superconducting fluctuations in the wire. These fluctuations may alter the numerical coefficient in S8 and hence the extraced value of $D_\varphi$.

## 2. Calibration of the persistent current signals
### 2.1 Estimation of persistent current from cantilever frequency

We infer the sample's persistent current $I$ from changes in the resonance frequency $\nu$ of the cantilever on which the sample is mounted. Here we describe the manner in which $I$ is inferred from $\nu$.

The cantilever's potential energy is a sum of two terms. The first is the elastic energy stored in the cantilever's deformation, which is very nearly a harmonic potential. The second term arises from the interaction between the ring's persistent current and the applied magnetic field. This magnetic term is a more complicated function of the cantilever's displacement, as both the persistent current and the resulting torque depend upon the angle between the ring and the applied field. We can write the magnetic energy as

$$E_{\text{mag}}(B,\theta) = \int^{B} I(B',\theta) A \sin\theta \, dB' \tag{S9}$$

where $A$ is the area of the ring and $\theta$ is the angle between the plane of the ring and the applied magnetic field $B$ (Fig 1A in the main paper). We note that a cantilever deflection leads to both a translation and a rotation of the sample; the latter changes $\theta$ and hence is responsible for the coupling between the persistent current and the cantilever.

We consider a persistent current which is approximately periodic[*] in the flux through the ring $\Phi = AB\sin\theta$. Expressed in terms of its Fourier components, the current is:

$$I(B) = \sum_{p} \left[ I^{(+)}_{h/pe} \sin(2\pi p \frac{BA\sin\theta}{\Phi_0}) + I^{(-)}_{h/pe} \cos(2\pi p \frac{BA\sin\theta}{\Phi_0}) \right] \tag{S10}$$

---

[*] This assumption of a persistent current which is periodic in the flux through the ring is a better approximation (for the purposes of this calculation) than might be gathered from the appearance of the persistent current data in the main paper (e.g., Fig. 1(e)), which clearly includes aperiodic modulations. This is because the data in the main paper is taken while varying the applied magnetic field, which adds flux both through the ring and the metal; the latter leads to aperiodic modulations of the current as discussed in the main paper. However when calculating the frequency shift at a *given* field, we are concerned with modulations of the flux due to the cantilever's motion. It is straightforward to show that this motion predominantly modulates the flux through the ring and not the flux through the metal.

Here $p$ is equivalent to the variable $f_\Phi$ in the main paper, and $p = 1$ corresponds to a period of $h/e$. The presence of magnetic flux inside the metal of the ring leads to aperiodic modulations of the PC oscillations; as a result we do not constrain $p$ to be an integer. This same modulation necessitates the cosine terms in Eq. S10, which are not present in treatments which assume time-reversal symmetry in the metal. We write the Fourier series as a sum over $p$ rather than an integral in order to correspond more closely with the data, which consists of discrete measurements of $I(B)$ and hence a discrete spectrum $\tilde{I}_{typ}(f_\Phi)$. Lastly, we note that the signal is expected (and observed) to be dominated by its Fourier components with $p \approx 1$. This fact will be used in some of the approximations below.

The magnetic energy $E_{mag}$ can then be evaluated from Eqs. S9 and S10:

$$E_{mag}^{(p)}(B,\theta) = \sum_p \frac{\Phi_0}{2\pi p}\left( I_{h/pe}^{(+)} \cos(2\pi p \frac{BA\sin\theta}{\Phi_0}) - I_{h/pe}^{(-)} \sin(2\pi p \frac{BA\sin\theta}{\Phi_0}) \right) \quad (S11)$$

The potential associated with the persistent currents (Eq. S11) is obviously not harmonic and hence will lead to a cantilever frequency shift which depends upon the amplitude of the cantilever's motion. For arbitrarily small amplitude, the frequency shift $\delta\nu$ is simply proportional to $\partial_\theta^2 E_{mag}$. However, this approximation is only valid when the cantilever's motion is small enough that the resulting modulation of the flux through the ring $\phi_{ac} \ll \phi_0$. In practice we typically use cantilever amplitudes which lead to $\phi_{ac} \sim \phi_0/2$, so we need to derive $\delta\nu$ for amplitudes relevant to our actual measurement. This calculation is done using canonical perturbation theory in the Hamilton-Jacobi framework, which is valid for $\delta\nu \ll \nu$, and hence appropriate for our measurements.

We begin by noting that the angle $\theta$ can be written as $\theta = \theta_0 + \gamma_m \frac{q_{tip}}{L}$ where $\theta_0$ is the value of $\theta$ at the cantilever's equilibrium position, $L$ is the cantilever's length, $q_{tip}$ is the displacement of the cantilever's tip from equilibrium, and $\gamma_m$ is the ratio between the slope of the cantilever at the position of the ring and the factor $q_{tip}/L$ for flexural mode $m$. For a ring located at the cantilever tip $\gamma_m = \{1.377, 4.788\}$ for the first two flexural modes.

The position of the cantilever can be rewritten in terms of the action-angle variables as $q_{tip} = \sqrt{\frac{2\nu}{k}J}\sin(2\pi\eta)$ where $k$ is the cantilever's mechanical spring constant, $J$ the action variable, and $\eta$ the canonically conjugate angle variable. The shift in cantilever resonant frequency is then given by[11]

$$\delta\nu = \partial_J \int_0^{2\pi} d\eta E(\theta(\eta, J)) \tag{S12}$$

$$= -\frac{AB\nu\gamma_m}{kLq_{max}}\cos(\theta_0)\sum_{p=1}^{\infty} J_1\left(2\pi p \frac{AB\cos(\theta_0)}{\Phi_0}\frac{\gamma_m q_{max}}{L}\right)\left(I_{h/pe}^{(+)}\cos\left(2\pi p \frac{\Phi}{\Phi_0}\right) - I_{h/pe}^{(-)}\sin\left(2\pi p \frac{\Phi}{\Phi_0}\right)\right)$$

where we have kept only the lowest-order term in the small parameter $\gamma_m \frac{q_{tip}}{L}$ and assumed that $\theta_0$ is chosen so that $\frac{\phi_0}{2\pi AB\tan\theta_0} \ll 1$. In Eq. S12, $J_1(x)$ is the first Bessel function of the first kind and $q_{max}$ is the amplitude of motion of the cantilever tip in units of distance. It can be readily verified that in the limit $q_{max} \to 0$ Eq. S12 reduces to $\delta\nu = \partial_\theta^2 E_{mag}$.

From this point we use two approaches to extract $I(B)$ from our measurement of $\delta\nu(B)$. In the first (Method A), we extract each Fourier component of $I(B)$ (i.e., the $p^{th}$) separately by converting the frequency shift data into the derivative $\frac{\partial I}{\partial B}$ using

$$\frac{\partial I_{h/pe}}{\partial B} \approx -\delta\nu \frac{2\pi pA\sin\theta_0}{\Phi_0}\left[AB\frac{\nu\gamma_m}{kLq_{max}}\cos(\theta_0)J_1\left(2\pi p\frac{AB\cos(\theta_0)}{\Phi_0}\frac{\gamma_m q_{max}}{L}\right)\right]^{-1} \tag{S13}$$

The approximation in going from Eq. S12 to Eq. S13 relies on the fact that the argument of the Bessel function varies only slightly over any given data set. Equation S13 is then integrated numerically with respect to $B$ to get $I_{h/pe}(B)$. $I_{h/pe}(B)$ is then Fourier transformed, and only the Fourier component at $f_\Phi = p$ is kept. This routine is repeated for each $p$, resulting in $\tilde{I}_\Sigma(f_\Phi)$; division by $\sqrt{N}$ gives $\tilde{I}_{typ}(f_\Phi)$, and inverse Fourier transform gives $I_{typ}(B)$.

The advantage of Method A is that it provides an accurate estimate of the Fourier components over a wide range of $p$ and so is suitable for data in which the persistent current has Fourier components near $h/e$ and $h/2e$ (we never observe a signal at higher $f_\Phi$). The disadvantage of Method A is that there are values of $p$ which lead to zeroes in the Bessel function in Eq. S.13, and any technical noise in the measurement of $\delta\nu$ will be converted to a diverging $I$ for these $p$ (equivalently, the measurement is not sensitive to these Fourier components of $I$). The values of $p$ at which these divergences occur can be modified by varying $q_{max}$, and we used this fact to verify for each sample that no signal (e.g., at $h/2e$) was being masked by this effect. Furthermore, for the samples with fairly large signals (the 308 nm and 418 nm arrays), $q_{max}$ was set low enough that the divergences in the rhs of Eq. S13 occurred at high values of $p$ (i.e., $> 4$) and so did not interfere with the inferred Fourier components at $h/e$ or $h/2e$. As a result, the $\tilde{I}_{typ}(f_\Phi)$ for these samples (Figs. 2E – F) was analyzed using Method A. $I_{typ}(B)$ for these samples (shown in Figs. 1E, 2A – B) is the inverse Fourier transform of $\tilde{I}_{typ}(f_\Phi)$, with the very lowest Fourier components (corresponding to the smooth background visible in the raw data, e.g., in Fig. 1D) discarded. The Fourier components where the rhs of Eq. S13 diverges were also discarded in calculating $I_{typ}(B)$.

For samples with weaker signals (the 418 nm ring and 793 nm array) $q_{max}$ could not be decreased far enough to push the divergences in Eq. S13 entirely out of the $f_\Phi$ band of interest (i.e., $0 < p < 2$). However by varying $q_{max}$ we confirmed that there was never a signal at $h/2e$ above the noise floor. To present the data from these samples we use a second method (Method B) which makes use of the fact that persistent current signal for these samples occurs entirely near $p \approx 1$ (i.e., has no observable components at $p \approx 2$) and that the argument of the Bessel function in Eq. S12 varies only weakly over a given data set. In this method we set $p = 1$ in the Bessel function and so rewrite Eq. S12 as

$$\delta\nu \approx -\frac{AB\nu\gamma_m}{kLq_{max}}\cos(\theta_0)J_1\left(2\pi\frac{AB\cos(\theta_0)}{\Phi_0}\frac{\gamma_m q_{max}}{L}\right)$$

$$\times \sum_{p=1}^{\infty}\left(I_{h/pe}^{(+)}\cos\left(2\pi p\frac{\Phi}{\Phi_0}\right) - I_{h/pe}^{(-)}\sin\left(2\pi p\frac{\Phi}{\Phi_0}\right)\right)$$

(S14)

This allows us to convert the measured frequency shift directly to $\frac{\partial I}{\partial B}$ (again taking $p \approx 1$):

$$\frac{\partial I}{\partial B} \approx -\delta\nu \frac{2\pi A \sin\theta_0}{\Phi_0} \left[ AB \frac{\nu\gamma_m}{kLq_{max}} \cos(\theta_0) J_1\left(2\pi \frac{AB\cos(\theta_0)}{\Phi_0} \frac{\gamma_m q_{max}}{L}\right) \right]^{-1} \quad (S15)$$

This quantity is then integrated numerically and Fourier transformed to get $\tilde{I}_\Sigma(f_\Phi)$ and then divided by $\sqrt{N}$ to get $\tilde{I}_{typ}(f_\Phi)$. Method B was used for the 418 nm ring and 793 nm array (Figs. 2G – H). It should be emphasized that the choice of $q_{max}$ for taking this data leads to an insensitivity to Fourier components of $I(B)$ corresponding to $h/2e$ period (i.e., the zeroes of Bessel function in Eq. S12); however by varying $q_{max}$ (and hence varying the location of the insensitive band) in other data runs we confirmed that there was no $h/2e$ periodic component of $I(B)$ above the noise floor.

The data showing $I_{typ}(B)$ for these samples (Figs. 2C – D) are the inverse Fourier transforms of $\tilde{I}_{typ}(f_\Phi)$ with the lowest Fourier components (corresponding to the smooth background visible in the raw data, e.g., in Fig. 1D) discarded. No high-frequency Fourier components were discarded.

Applying Methods A and B to a single data set typically results in values of $\langle I^2_{h/pe} \rangle^{1/2}_D$ which differ by ~ 4%. The effect of this uncertainty is discussed below in Section 2.3.

**2.2 Tests of the conversion between frequency shifts and persistent currents**

We tested the accuracy of Eqs. S12 – S15 in several ways. A direct comparison between Eq. S13 and a measurement of $\delta\nu$ versus $q_{max}$ at a single value of $B$ is difficult to perform, due to the fact that the cantilever frequency depends on $q_{max}$ even in the absence of a magnetic field (presumably because of the weak intrinsic nonlinearity of the cantilever's mechanical properties). To remove this nonmagnetic background, we measured $\delta\nu$ as a function of $q_{max}$ at two similar magnetic fields, as indicated in Figure S6A. After subtracting the two curves (shown separately in the inset of Fig. S6B) the remaining frequency shift is dominated by the component due to the persistent current. The result is shown in the main body of Fig. S6B, along with a fit to equation S12 (with $p = 1$) with reasonable parameters for the ring's size and the amplitude of the $h/e$ component of the current.

Figs. S7-S8 show two other tests, in which the inferred current was observed to be unaffected over a wide range of cantilever amplitudes and for excitation of two different cantilever flexural modes.

The quality of the fit in Fig. S6, along with the data in Figs. S7-S8, implies that our method of extracting the persistent current from the frequency shift is accurate and, furthermore, that the cantilever's motion in the magnetic field does not produce any appreciable nonequilibrium effects.

**2.3 Uncertainty estimates**

Uncertainties in the measurements of $\langle I^2_{h/pe} \rangle_D^{1/2}$ shown in Fig. 3 of the main paper arise from a number of sources which we list here.

1) We estimate that our temperature measurements have an uncertainty of 7% (based on the manufacturer's specifications and comparison with fixed points).

2) The statistical error in our estimation of the background $b(f_\Phi)$ ranges from 0.5 pA (for the 793 nm array) to 10 pA (for the 418 nm ring).

3) Since the quantity $\langle I^2_{h/pe} \rangle_D^{1/2}$ is itself a variance of a distribution that we are estimating from a finite data set, the uncertainty in our estimate will be given by the Standard Error of the Variance (SEV). The SEV is related to the number of independent realizations (in our case, this number is the ratio between the span of $B$ over which the measurement is made and $B_c$). For our data the SEV error in $\langle I^2_{h/pe} \rangle_D^{1/2}$ ranges from 6% (for the 418 nm array) to 20% (for the 408 nm ring).

4) We estimate that the uncertainty introduced by the various approximations in Methods A and B is 4%.

Each of these sources of uncertainty will result in errors with varying degrees of correlation between the individual data points in Fig. 3. For example the error due to the SEV should be constant for a given sample (cf. Fig. S18) and hence lead to an unknown (but constant) scaling of the data from each sample, while the error in $b(f_\Phi)$ should be random for each measurement, leading to scatter in Fig. 3. The uncertainties in $D$ quoted in Table 1 correspond to the statistical error in the fits of Fig. 3, and hence reflect the sources of error leading to scatter in Fig. 3. Based on standard error propagation, we estimate that uncertainty in the overall scaling of the curves in Fig. 3 leads to an additional uncertainty ~ 6% in each value of $D$.

### 3. Measurement diagnostics

We performed a variety of diagnostic measurements to characterize the invasiveness of our cantilever-based detector. In particular we wanted to ensure that the measurement does not induce spurious non-equilibrium effects that may obscure or mimic the PC signal. The effects of cantilever oscillation amplitude and frequency are discussed in the Section 2.2; here we show the effects of readout laser power, magnetic field polarity, magnet persistent mode, and the presence of room-temperature electronics connected to the cryostat.

Fig. S9 shows the persistent current as a function of $B$ for five different laser powers $P_{inc}$, where the laser is used for interferometric detection of the cantilever's position. The signal is independent of laser power between $P_{inc}$ = 0.8 nW and 80 nW and is only slightly affected at $P_{inc}$ = 800 nW. The data in the main paper were taken with $P_{inc}$ ~ 5 nW. In particular, the data in Fig. S9 indicates the absence of heating at the laser powers used in the experiment.

We measured the persistent current at both positive and negative magnetic fields and the results are plotted in Fig. S10, with the $x$-axis negated for the negative $B$ trace. The signal is unaffected by the polarity of the field.

Lastly, we took data with the magnet in different operational modes: (1) the magnet persisted and the current in the magnet leads ramped down to zero, (2) the magnet persisted with current flowing in the magnet leads, and (3) the magnet not persisted. To minimize RF electronic noise in mode (1), all electronics were disconnected from the dewar (e.g., thermometry, heaters, liquid helium level meters, etc.) except the PZT drive which was fed through a room-temperature 1.9 MHz coaxial low pass filter. The results plotted in Fig. S11 show no dependence of the persistent current signal on the operating mode of the magnet.

### 4. Magnetic Field Sweeps

As discussed in the main paper, measuring the persistent current over a range of $B$ spanning many $B_c$ allows us to determine the disorder averaged current $\left\langle I_{h/e}^2 \right\rangle_D^{1/2}$. Figures S12-S17 show data from these large $B$ sweeps for each of the samples measured. In each of these figures the quantity plotted is $I'(B)$, where

$$I'(B) = \frac{\Phi_0}{2\pi A \sin\theta} \frac{\partial I}{\partial B} \qquad (S16)$$

which is derived from measurements of the cantilever frequency shift using Eq. S15. The choice of normalization in Eq. S16 means that the oscillations of $I'(B)$ in Figs. S12 – S17 have the same amplitude as the oscillations in $I(B)$.

These large sweeps provide a direct measure of $\langle I_{h/e}^2 \rangle_D^{1/2}$ (the quantity relevant for making comparisons with theory), and result in the points marked by arrows in Fig. 3 of the main paper. For the remainder of the points in Fig. 3, we make use of the fact that $I(B)$ is found empirically to depend upon $T$ only via an overall scaling, as shown in Fig. S18. To generate the points in Fig. 3 not marked by arrows, this scaling is applied to the value of $\langle I_{h/e}^2 \rangle_D^{1/2}$ determined from the large $B$ sweeps.

## 5. Supplementary Figures

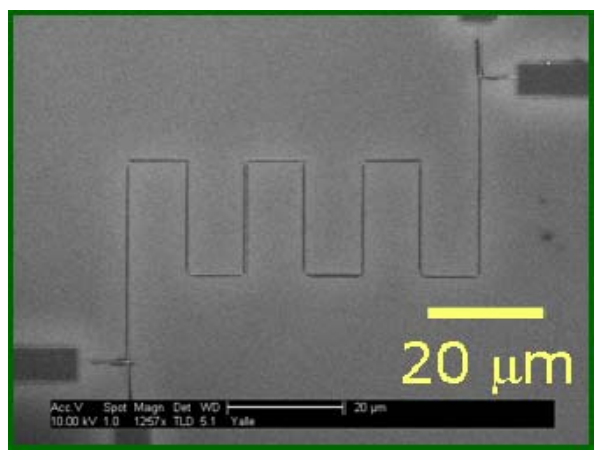

**Figure S1** SEM image of a 289 μm long wire similar to the one used for transport measurements.

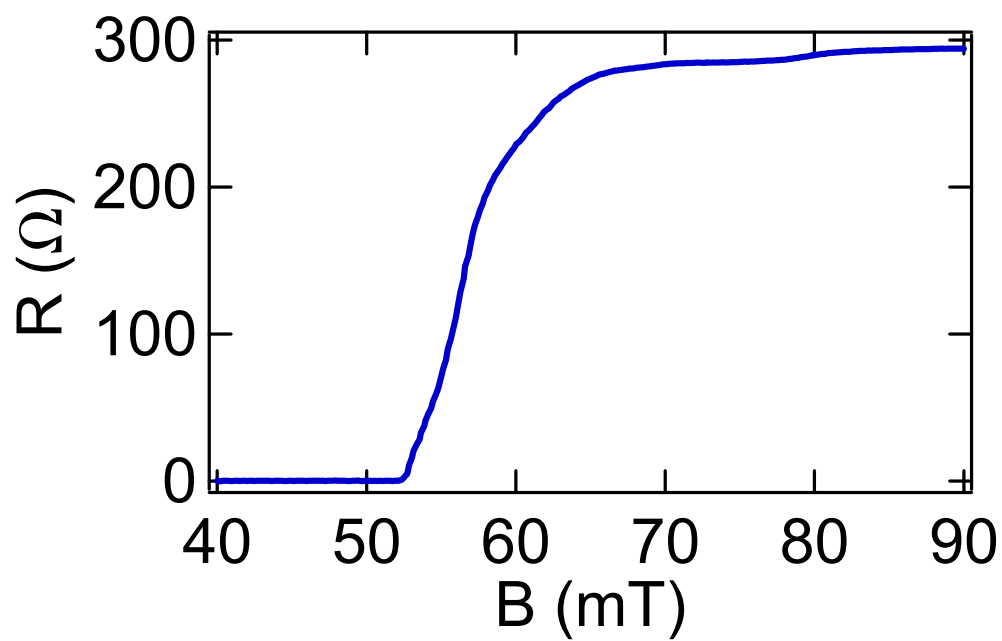

**Figure S2** Resistance versus magnetic field at 365 mK for the transport wire. The superconducting transition occurs over a range of magnetic field beginning at 52 mT. The magnetic field was swept in the direction of increasing field strength.

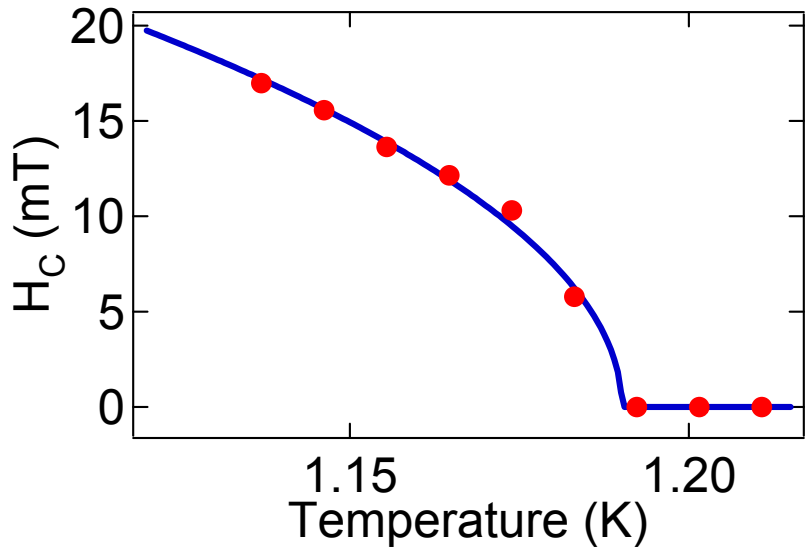

**Figure S3** Superconducting critical field versus temperature for the transport wire. For the data shown, $H_c(T)$ was taken to be the field at which the resistance was ten percent of the normal state value. The data are fit using Eq. S2 with $D_{Hc}$ and $T_c$ as fitting parameters. Defining $H_c(T)$ to occur at a different percentage of the normal state resistance value shifts the fitted $T_c$ slightly but does not affect $D_{Hc}$.

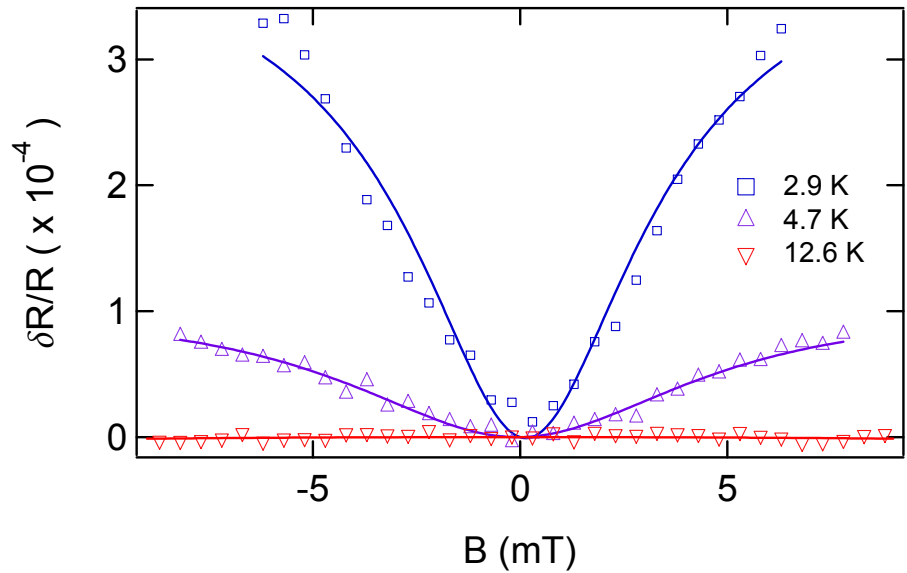

**Figure S4** Change in resistance versus magnetic field with fits to Eqs. S3 and S6 for three different temperatures. In order to achieve an adequate signal to noise ratio, the data were fit over ranges shown despite the conditions cited in the text for validity of Eq. S6.[8]

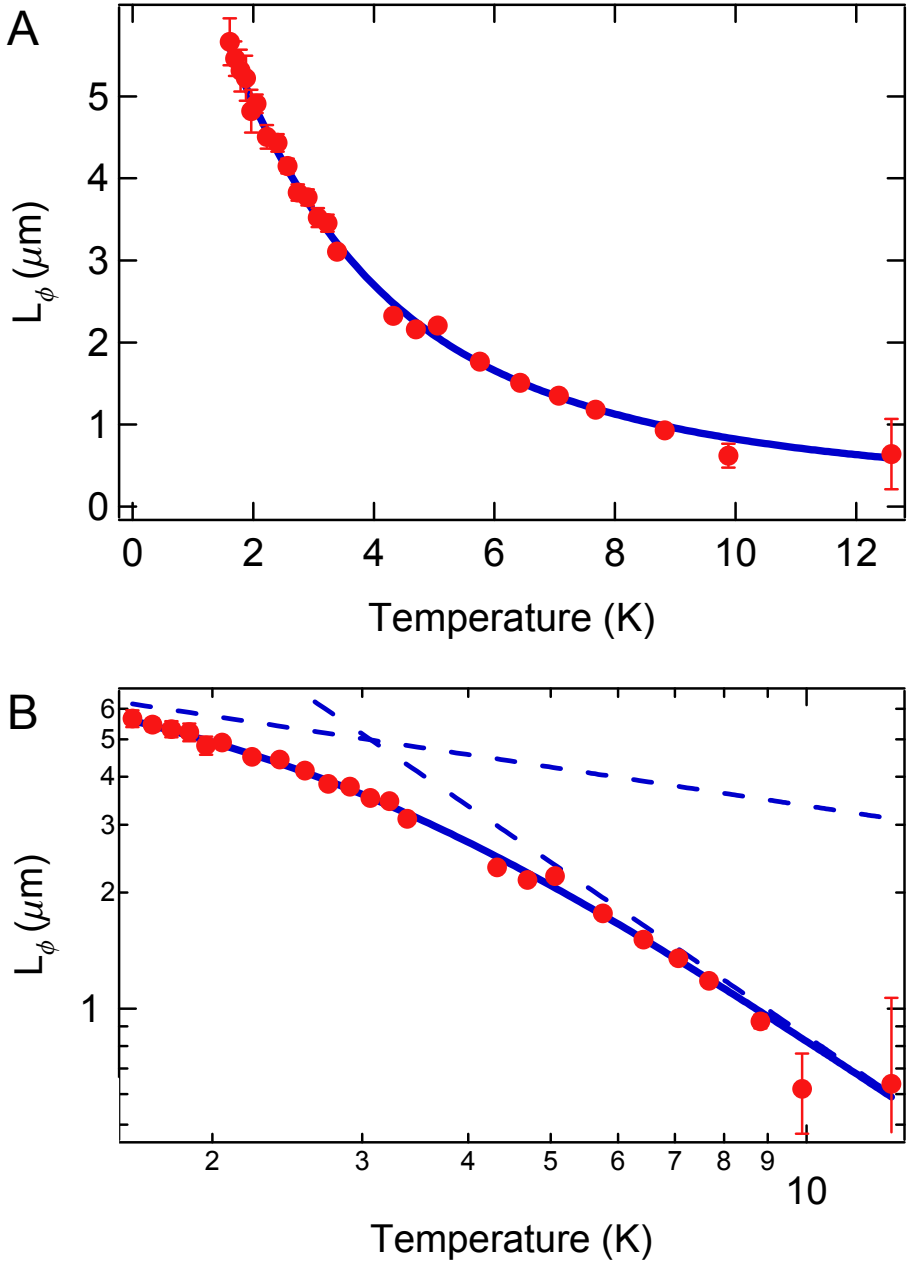

**Figure S5** Electron phase coherence length $L_\phi$ versus temperature. The dots represents values extracted from fits to the magnetoresistance of the 289 μm wire. The same data is plotted on a linear scale (A) and a log-log scale (B). The solid line is the fit to the functional form described in the text, and the dashed lines indicate the specific contributions to this fit from electron-electron scattering and electron-phonon scattering, as discussed in the text.

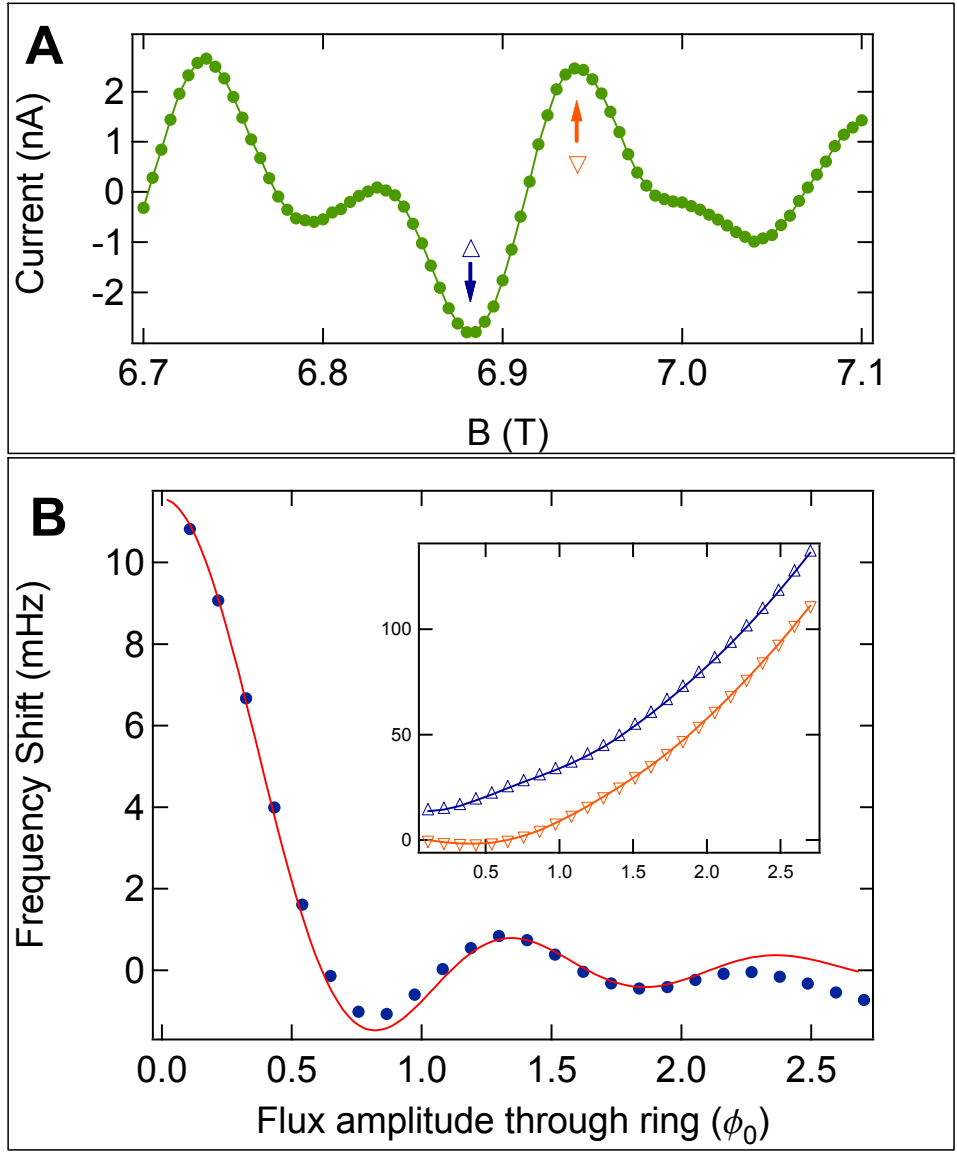

**Figure S6 (A)** Detail of the persistent current versus magnetic field for the array of 308 nm rings. For the measurement shown, the cantilever was oriented with $\theta = 6°$. The arrows indicate two field values at which measurements of the cantilever frequency shift were performed as a function of cantilever amplitude. **(B)** Difference in cantilever frequency shift for the two field values indicated in **(A)** versus cantilever amplitude of oscillation. The cantilever was driven in its second flexural mode, which had a frequency of 13718 Hz and a spring constant of 0.053 N/m. The cantilever had a length of 449 μm. The cantilever amplitude is plotted on the x-axis in terms of the amplitude of the flux modulation $\phi_{ac}$ (in units of the flux quantum) through the ring produced by the cantilever motion. The solid curve is a fit using Eq. S12 (with $p = 1$) with $I_1 = 5.5 \pm 0.2$ nA and $r = 265 \pm 2$ nm. The inset shows the frequency shift measured at the two points indicated in **(A)** and has the same units as the main plot in **(B)**.

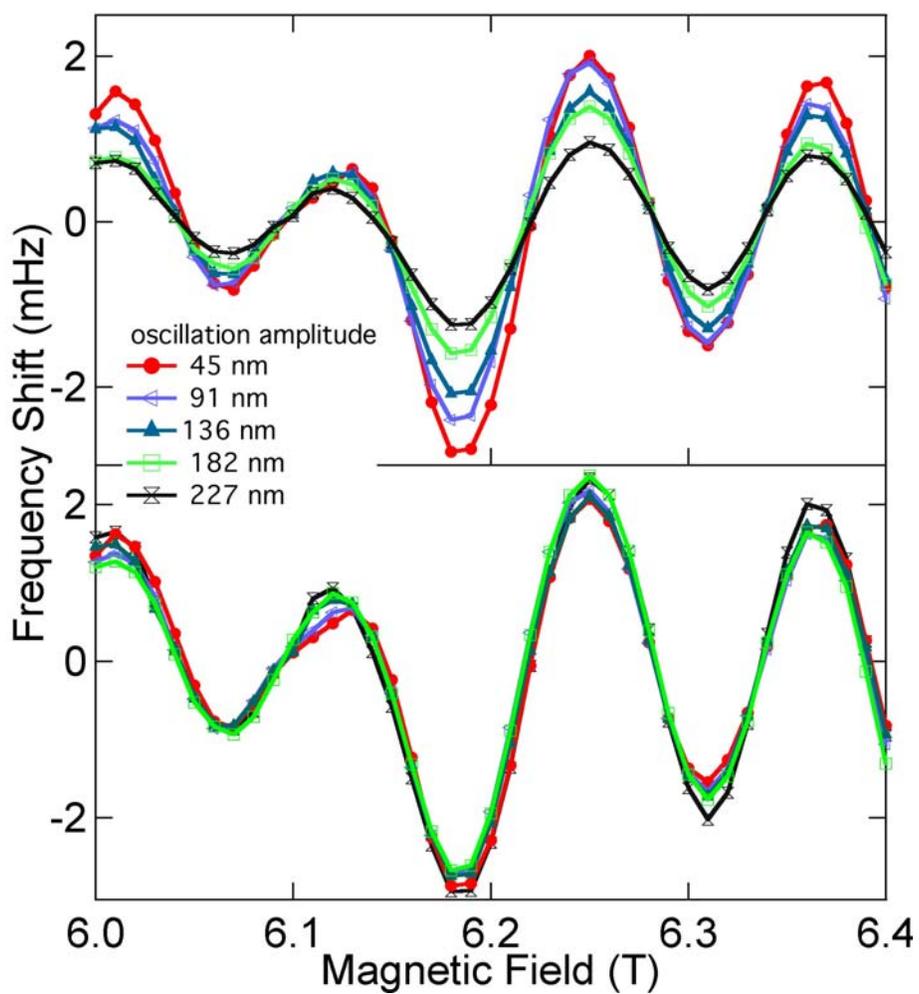

**Figure S7** Cantilever frequency shift versus magnetic field for five different cantilever oscillation amplitudes $q_{max}$ (measured at the location of the rings). The lower panel shows the same data as the upper panel, but scaled so as to give $\delta\nu(q_{max}=0)$ using Eq. S12 with $p=1$. The traces collapse on top of each other, indicating that they are due to equilibrium persistent currents.

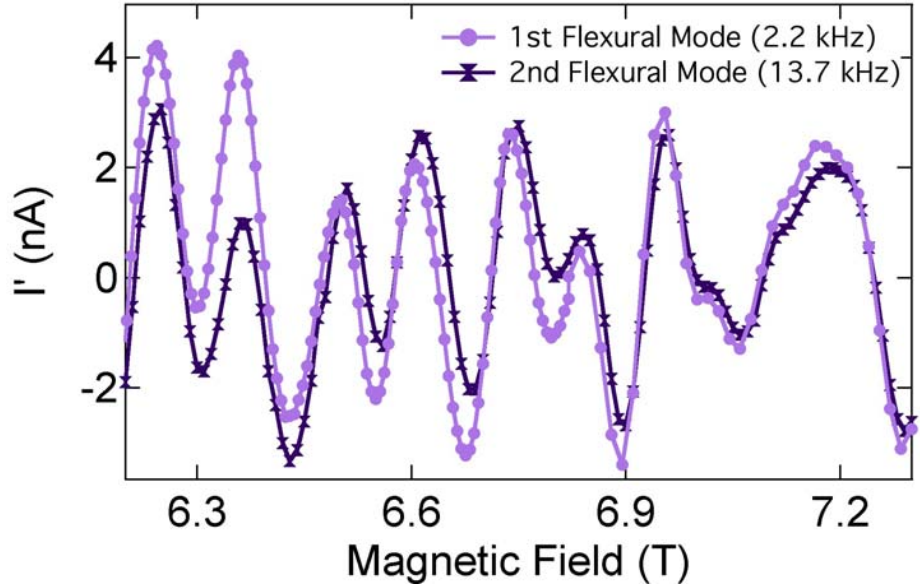

**Figure S8** The derivative of the persistent current *I′* (derived from Eqs. S15 and S16) versus magnetic field measured when oscillating the cantilever at 2.2 kHz (the cantilever's first flexural resonance) and at 13.7 kHz (the cantilever's second flexural resonance). The persistent current does not depend on the cantilever oscillation frequency; the slight difference in the two curves' smooth backgrounds is presumably due to different mechanical resonances present in the sample holder at 2.2 and 13.7 kHz.

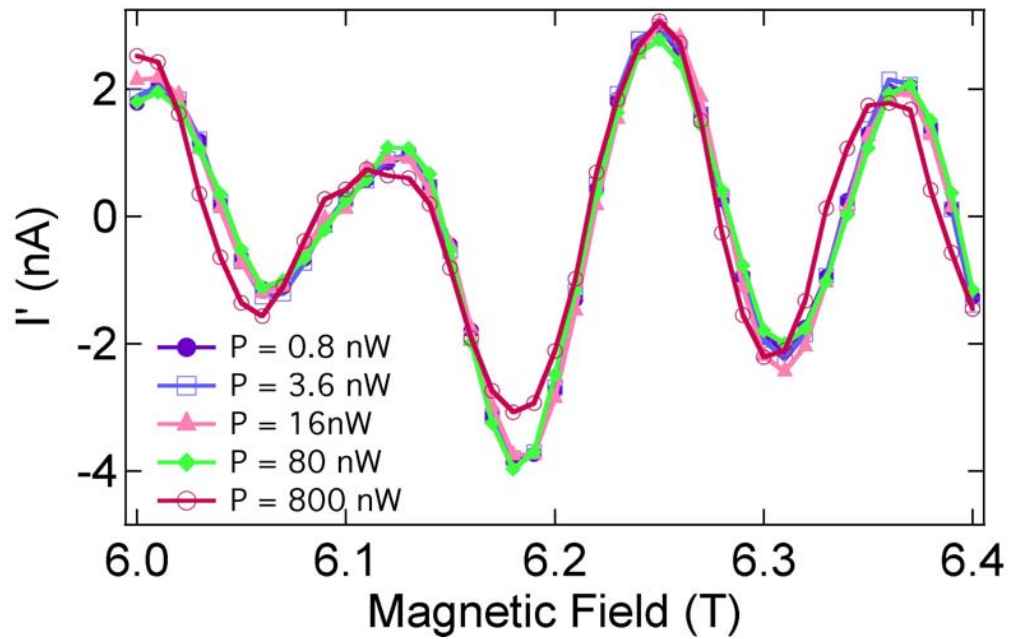

**Figure S9** The derivative of the persistent current *I′* (derived from Eqs. S15 and S16) versus magnetic field for a series of laser powers incident on the cantilever. For the data shown in the main paper, 5 nW of laser power was used.

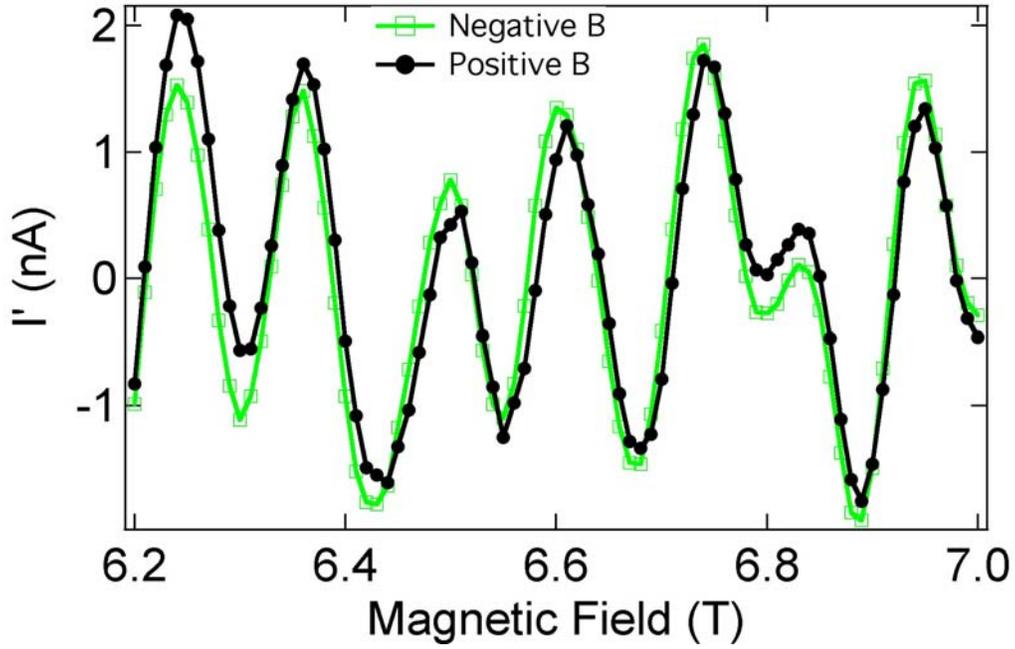

**Figure S10** The derivative of the persistent current $I'$ (derived from Eqs. S15 and S16) versus magnetic field for both magnetic field polarities. The x-axis of the negative magnetic field trace is negated.

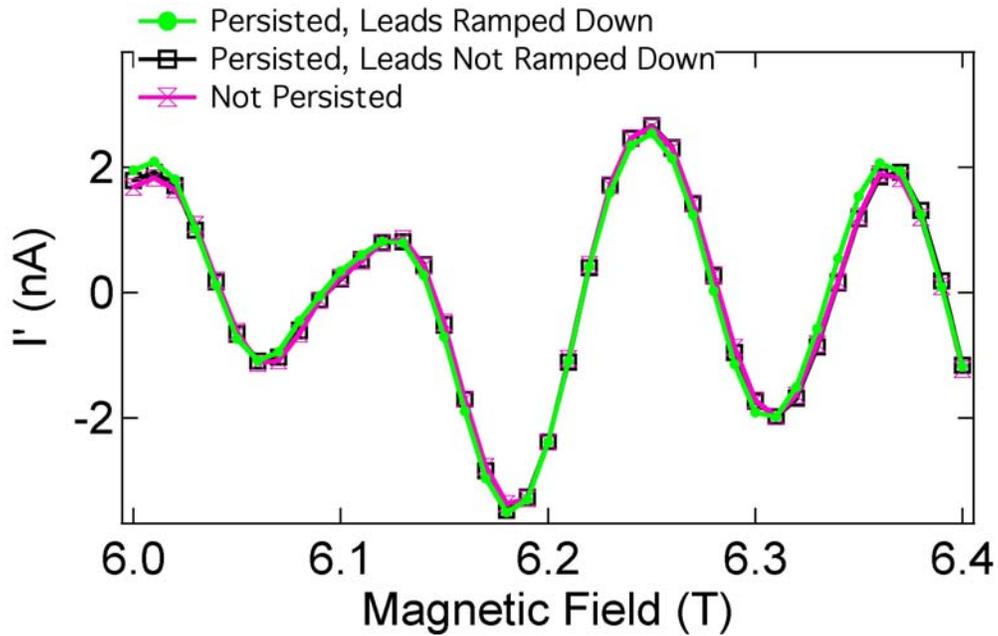

**Figure S11** The derivative of the persistent current $I'$ (derived from Eqs. S15 and S16) versus magnetic field under different modes of operation of the superconducting solenoid producing the magnetic field. The operational modes of the magnet are indicated in the figure. For the data shown as green dots, all electronics (except for the piezo drive) were disconnected from the cryostat.

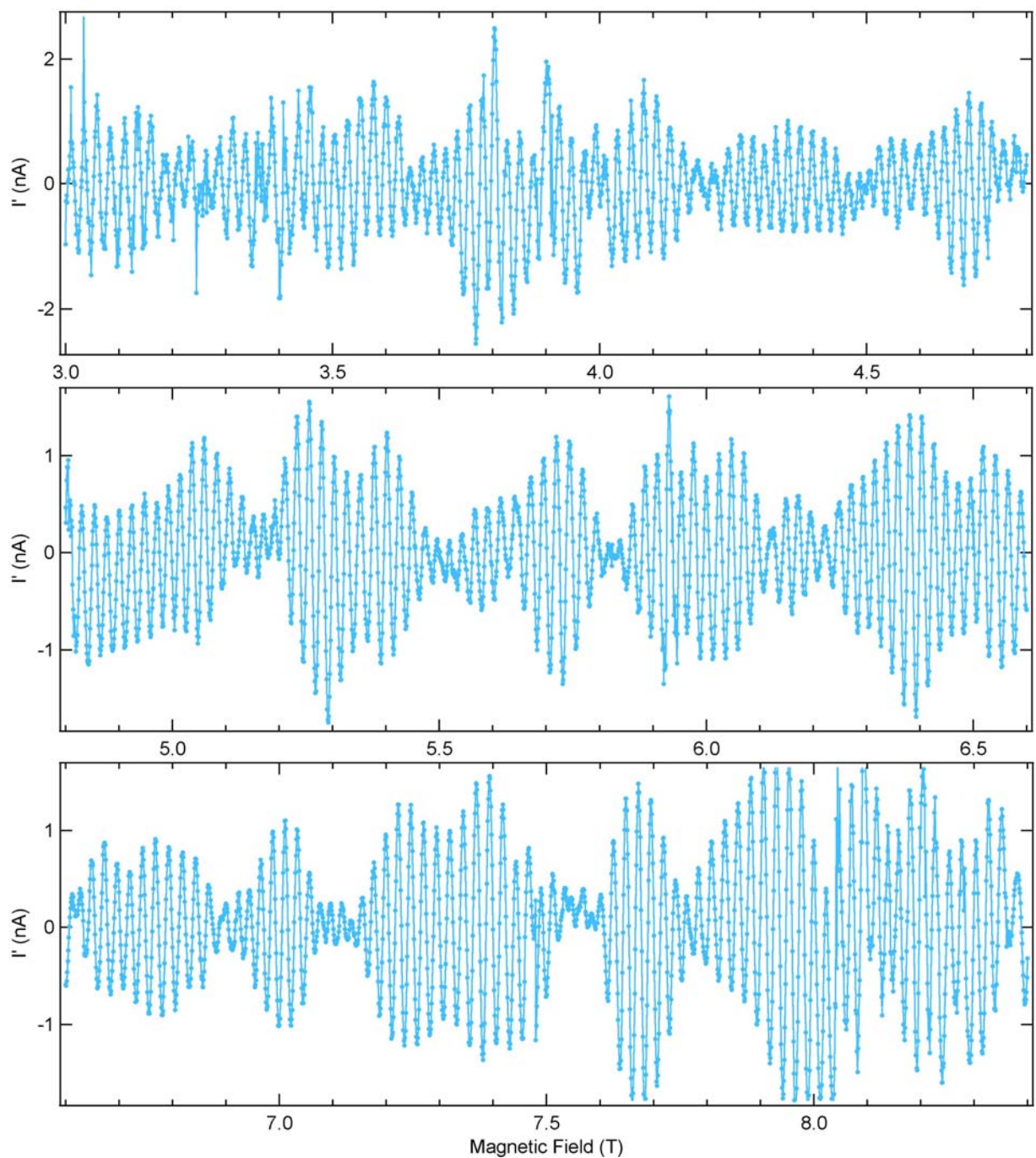

**Figure S12** The derivative of the persistent current $I'$ (derived from Eqs. S15 and S16) versus magnetic field for an array of 1680 rings with radius 308 nm at $T = 365$ mK. The full sweep is separated into three panels for clarity. The field is applied at 45° to the plane of the rings.

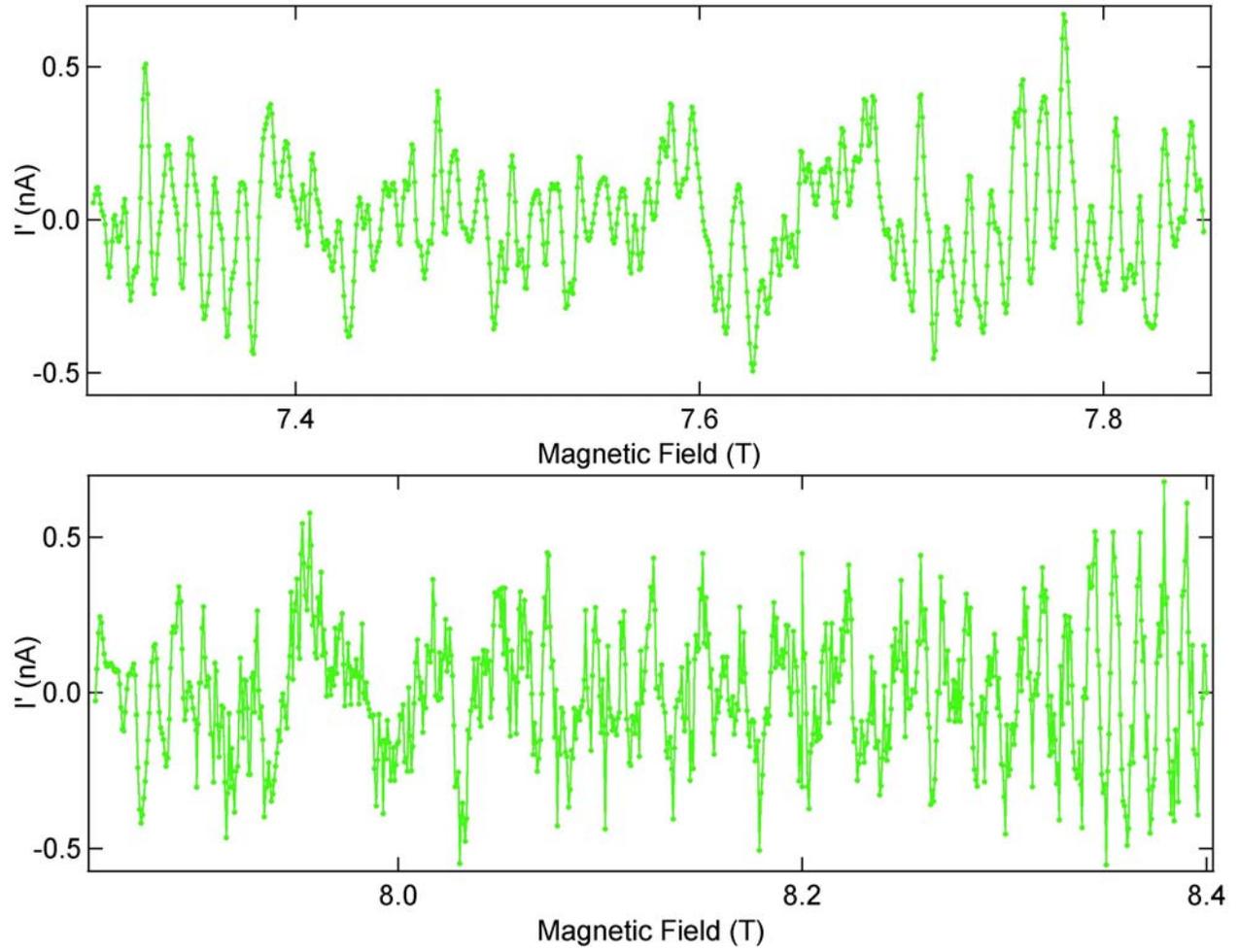

**Figure S13** The derivative of the persistent current $I'$ (derived from Eqs. S15 and S16) versus magnetic field for a single ring of radius 418 nm at $T = 365$ mK. The field is applied at 45° to the plane of the ring. The full sweep is separated into two contiguous panels for clarity.

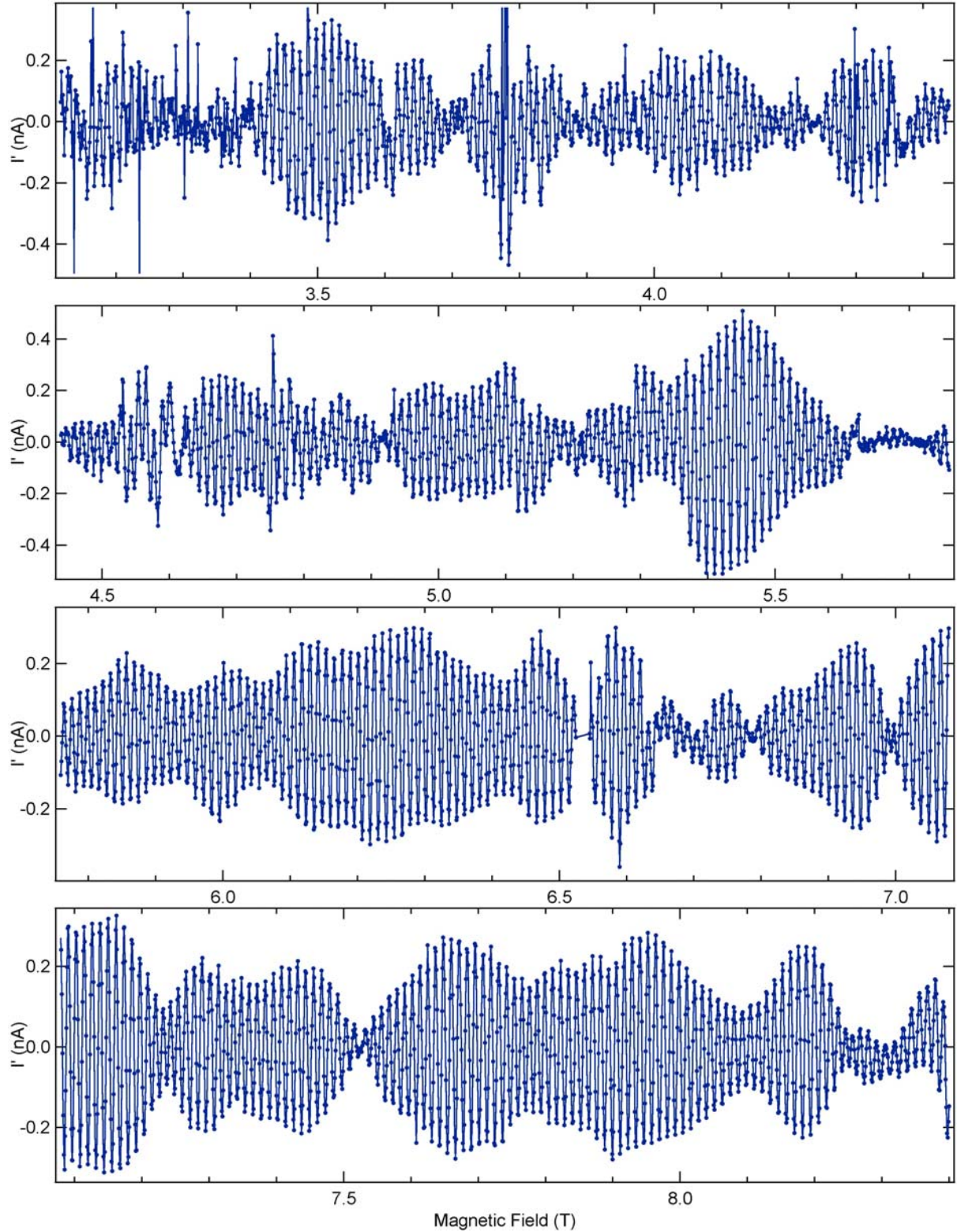

**Figure S14** The derivative of the persistent current $I'$ (derived from Eqs. S15 and S16) for an array of 990 rings with $r = 418$ nm at $T = 365$ mK. and $\theta = 45°$ to the plane of the rings.

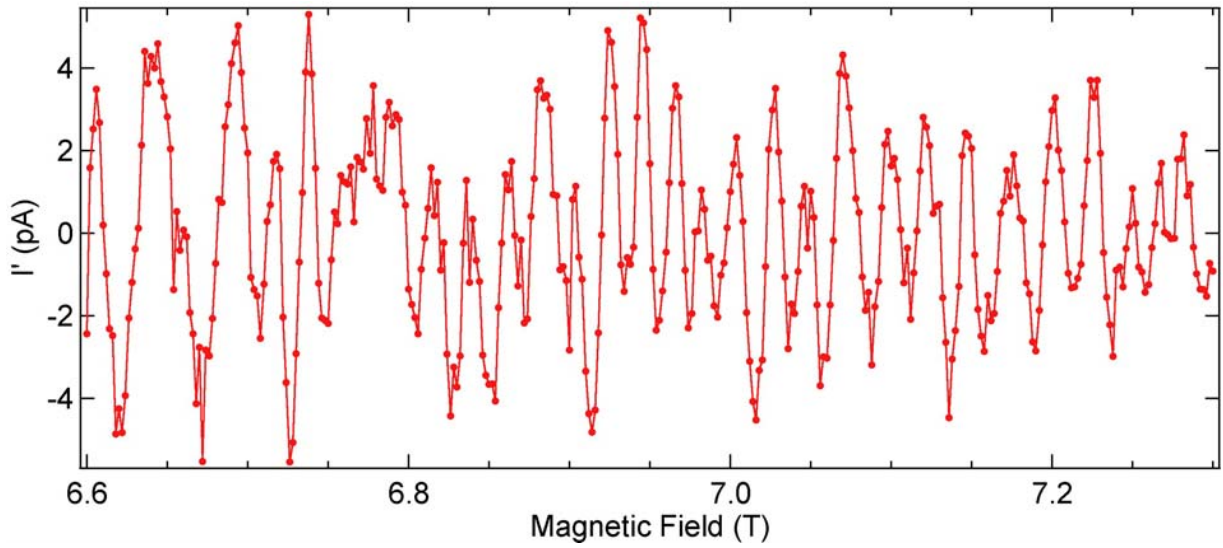

**Figure S15** The derivative of the persistent current $I'$ (derived from Eqs. S15 and S16) versus magnetic field for an array of 282 rings with radius 793 nm at $T = 323$ mK. The field is applied at 6° to the plane of the rings.

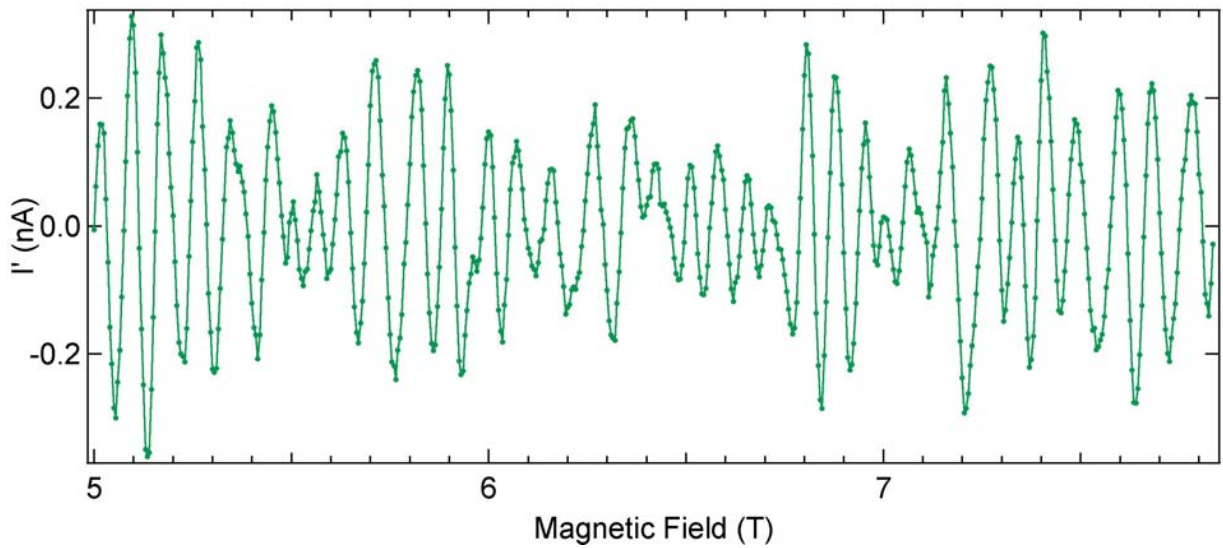

**Figure S16** The derivative of the persistent current $I'$ (derived from Eqs. S15 and S16) versus magnetic field for an array of 990 rings with radius 418 nm at $T = 323$ mK. The field is applied at 6° to the plane of the ring.

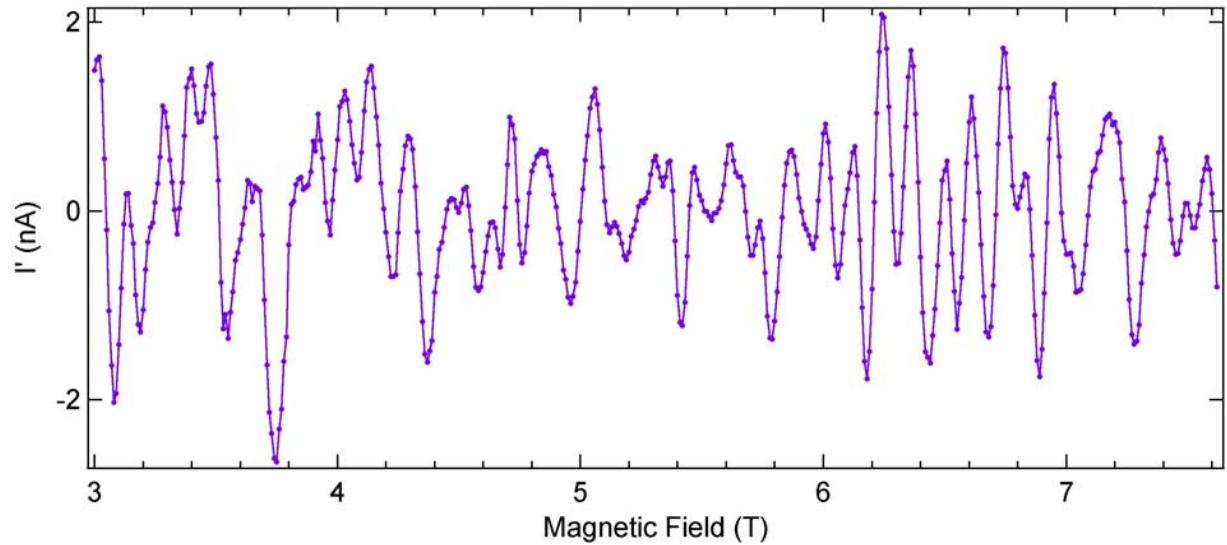

**Figure S17** The derivative of the persistent current $I'$ (derived from Eqs. S15 and S16) versus magnetic field for an array of 1680 rings with radius 308 nm at $T$ = 323 mK. The field is applied at 6° to the plane of the ring.

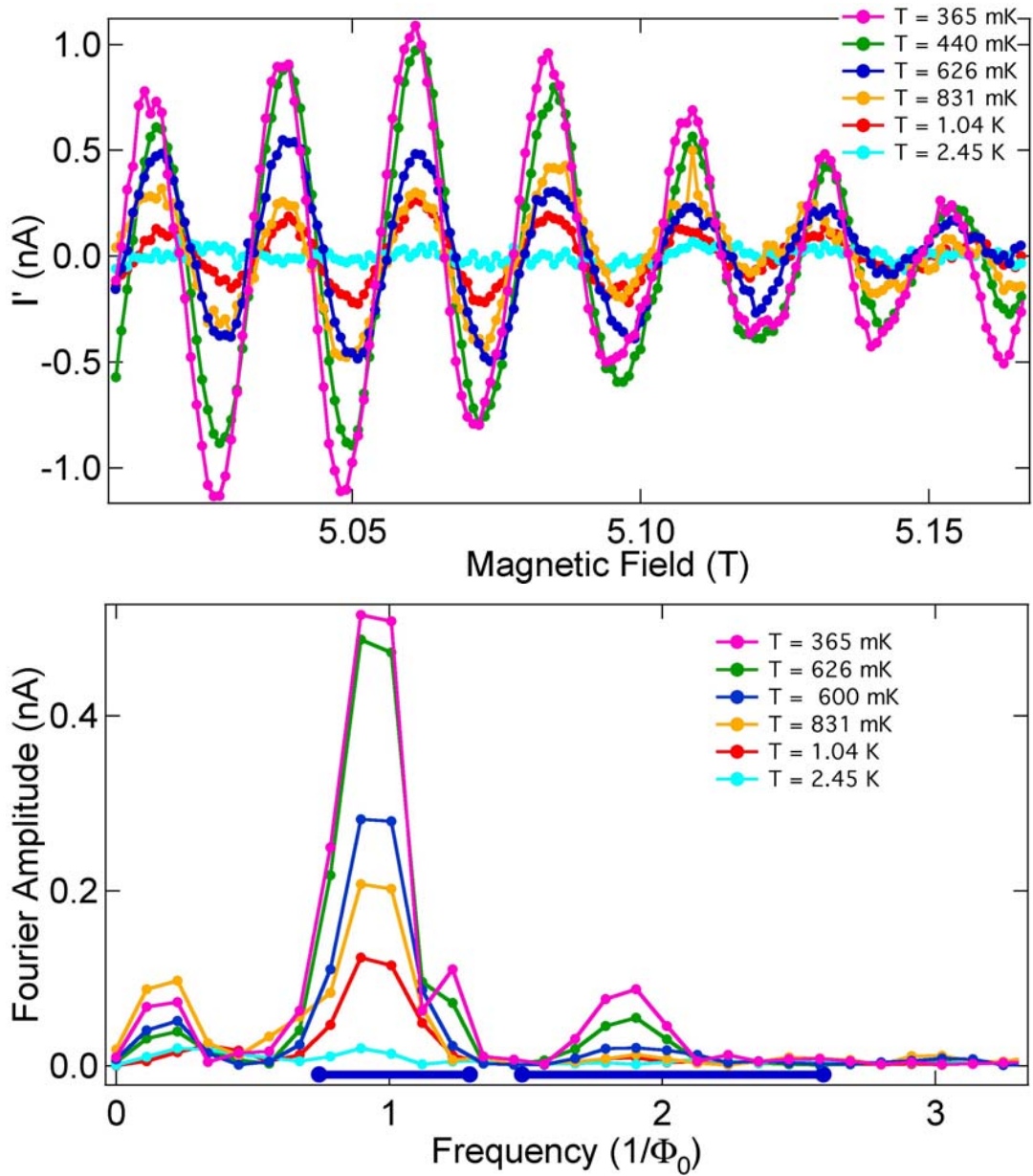

**Figure S18** The derivative of the persistent current $I'$ (derived from Eqs. S15 and S16) versus magnetic field (upper plot) for the array of $r$ = 308 nm rings measured with the field applied at 45° to the plane of the rings. The lower plot shows the Fourier transform of the same data. Traces are taken at different temperatures, as indicated in the figure legend. The amplitude of current oscillations decreases with increasing temperature, but the shape of the oscillations remains unchanged.